\documentclass[12pt]{iopart}

\usepackage{bm}
\usepackage{graphicx}
\usepackage{amssymb}

\usepackage[usenames]{color}

\newcommand{\beq}{\begin{equation}}
\newcommand{\beqa}{\begin{eqnarray}}
\newcommand{\eeq}{\end{equation}}
\newcommand{\eeqa}{\end{eqnarray}}
\newcommand{\et}{{\it et al.}}

\newcommand{\lmk}{\left(}
\newcommand{\rmk}{\right)}

\newcommand{\lla}{\left\langle}
\newcommand{\p}{\partial}
\newcommand{\rra}{\right\rangle}
\newcommand{\so}{M_\odot}
\newcommand{\mch}{{\cal M}}

\newcommand{\mrm}{\mathrm }

\begin{document}

\title[GW observations of galactic IMBH binaries with DPF]{Gravitational wave observations of \\ 
galactic intermediate-mass black hole binaries \\
 with DECIGO Path Finder}

\author{Kent Yagi}

\address{ 
Department of Physics, Kyoto University,
   Kyoto, 606--8502, Japan}
\ead{kent@tap.scphys.kyoto-u.ac.jp}

\begin{abstract}

DECIGO Path Finder (DPF) is a space-borne gravitational wave (GW) detector
with sensitivity in the frequency band 0.1--100Hz. As a first step mission to DECIGO, it is
aiming for launching in 2016--2017. Although its main objective is to demonstrate technology
for GW observation in space, DPF still has a chance of detecting GW
signals and performing astrophysical observations. With an observable range up to 50
kpc, its main targets are GW signals from galactic intermediate mass black hole
(IMBH) binaries. By using inspiral-merger-ringdown phenomenological waveforms, we
perform both pattern-averaged analysis and Monte Carlo simulations including the
effect of detector motion to find that the masses and (effective) spins of the IMBHs could
be determined with errors of a few percent, should the signals be detected. Since GW signals
from IMBH binaries with masses above $10^4 M_\odot$ cannot be detected by ground-based
detectors, these objects can be unique sources for DPF. If the inspiral signal of a $10^3M_\odot$ 
IMBH binary is detected with DPF, it can give alert to the ringdown signal
for the ground-based detectors $10^2$--$10^3$s before coalescence. We also estimate the
possible bound on the graviton Compton wavelength from a possible IMBH binary in
$\omega$ Centauri. We obtain a slightly weaker constraint than the solar system experiment
and an about 2 orders of magnitude stronger constraint than the one from binary pulsar
tests. Unfortunately, the detection rate of IMBH binaries is rather small.  

\end{abstract}

\maketitle

\section{Introduction}

There are several ground-based interferometers aiming for the first direct detection of gravitational waves (GWs) from astrophysical sources such as inspirals of neutron-stars (NSs), black holes (BHs) and mixed binaries, and also from supernovae and gamma-ray bursts (GRBs).
Among them, LIGO is now under upgrading phase to advanced LIGO~\cite{ligo1,ligo2}.
VIRGO is currently operating, but will soon become offline and will be upgraded to advanced VIRGO~\cite{virgo1,virgo2}, while GEO~\cite{geo1,geo2} will continue to take data. 
TAMA is also in upgrading phase to LCGT~\cite{lcgt1,lcgt2}.
%
These detectors have their best sensitivity around 100--1000Hz and their lower frequency sides are limited by the seismic noises\footnote{Initial LIGO had a cutoff frequency at 40Hz.}.
To overcome this limit, space-borne interferometers have been proposed.
Among them, the Laser Interferometer Space Antenna (LISA) has been proposed with its optimal sensitivity around the GW frequency of 1mHz~\cite{danzmann,lisa}.
The expected targets are supermassive black hole (SMBH) binaries and white dwarf (WD) binaries.
Unfortunately, the latter GWs may cover up other signals including a primordial GW background~\cite{allen, maggiore}.
As a prototype mission, LISA path finder (LPF) will be launched before LISA~\cite{lpf}.

Another space-borne interferometer, the Deci-Hertz Interferometer Gravitational Wave Observatory (DECIGO)~\cite{decigo,kawamura2008} has been proposed in Japan.
It is most sensitive at around 0.1--1Hz.
Similar interferometer, the Big Bang Observatory (BBO) has been suggested as a follow-on mission to LISA~\cite{phinneybbo,cutlerharms,cutlerholz}. 
Thanks to the  high frequency cutoffs of WD/WD binary signals at $f\sim 0.2$Hz~\cite{farmer}, the main target of DECIGO and BBO is the primordial GW background (PGWB).
There is a possibility of NS/NS foreground signals masking PGWB but it has been shown that BBO is expected to have enough sensitivity to subtract sufficient amount~\cite{cutlerharms,harms}.
DECIGO may have to improve its sensitivity by 2--2.5 times in order to achieve this goal~\cite{yagiseto}. 
They have other interesting scientific objectives including direct detection of the accelerating expansion of the universe~\cite{decigo}, performing the precision cosmology~\cite{cutlerholz}, revealing the thermal history of the early universe~\cite{nakayama1,nakayama2} and the formation process of SMBH~\cite{gair}, verifying the primordial BH as the dark matter candidate~\cite{saito1,saito2}, and probing alternative theories of gravity~\cite{kent2} and the size of extra dimension~\cite{kentbrane}.

DECIGO Path Finder (DPF) is planned as a first step mission for DECIGO, hopefully launched in 2016--2017~\cite{ando1,ando2}.
Its main goals are to test the key technologies for the space mission and to perform observations of GWs and Earth gravity. 
As for the first space-borne GW detector, the torsion-bar type space antenna called SWIM has already been launched in 2009
 and successfully performed its observation run~\cite{toba}.
It has completed its mission and is currently under the phase of data analysis.
Recently, the improved version of the torsion-bar antenna (TOBA) has been proposed by Ando \textit{et al}.~\cite{toba}.
In this paper, we focus on the scientific significances of GW observations with DPF.

The main source for DPF is the intermediate mass BH (IMBH) binaries, where an IMBH refers to a BH having a mass of $10^2$--$10^4 M_\odot$.
(See e.g. Refs.~\cite{pau1,pau2} for the studies of detecting GWs from IMBH binaries with LISA.)
There is plenty of evidence that there exist stellar-mass BHs and SMBHs, but there is no direct detection of individual IMBHs and their existence is still disputed.
One of the possible evidence for their existence is the discovery of the ultraluminous X-ray sources (ULXs)~\cite{colbert} whose luminosity exceeds the Eddington luminosity of a stellar-mass BH, though their dynamical friction timescales are longer than the ones for SMBHs.  
Other possible evidence includes the radial mass-to-light ratio~\cite{gerssen} and the surface brightness profiles~\cite{noyola,miocchi} of globular clusters, stellar proper motions near their centers~\cite{anderson1,anderson2} and observations of several millisecond pulsars in the galactic globular cluster~\cite{ferraro}.
IMBH detection is important since it will give us clues for the formation mechanism of SMBHs~\cite{van}.
Since the observable range of DPF for these sources are within 50kpc, DPF targets are IMBH binaries in our Galaxy.
There are possibilities that IMBHs exist at the center of globular clusters and massive young clusters (see Refs.~\cite{colbert,pasquato} for reviews on IMBHs) and they may form binaries~\cite{gurkan,amaro}.

We perform Fisher analysis using the spin-aligned phenomenological inspiral-merger-ringdown waveform~\cite{ajith-spin} to estimate how accurately we can measure binary parameters if the signal has been detected.
We perform both pattern-averaged analysis and Monte Carlo simulations, taking the motion of DPF into account for the latter.
We also comment on the possible joint search with DPF and ground-based detectors.
DPF may be able to give alert to the latter about 10 mins before the coalescence.
Also if only the ringdown signal is obtained with the ground-based interferometers, DPF data including inspiral and merger information may help in distinguishing the signal from noises.
Furthermore, by combining the data of DPF and the ground-based detectors, the ringdown efficiency $\epsilon_\mrm{rd}$~\cite{flanaganhughes,bcw} can be determined, which cannot be measured from the ringdown signal alone due to the degeneracy against the distance $D_L$ to the source.

Following Ref.~\cite{keppel}, we also consider the possible constraint on the graviton Compton wavelength $\lambda_g$ from the DPF observations of IMBH binaries in our galaxy (see Refs.~\cite{rubakov,hinterbichler} for the reviews on massive gravity theories).
We compare our result with the current constraints on $\lambda_g$, (i) $\lambda_g \geq 2.8\times10^{17}$cm~\cite{talmadge} from the solar system experiment and (ii) $\lambda_g \geq 1.6\times10^{15}$cm~\cite{sutton} from binary pulsar observations.
DPF constraint would be important since it is the bound in the strong-field regime, whereas the current constraints mentioned above have been obtained both in the weak-field regime.

Unfortunately, the expected detection rate of IMBH binaries with DPF is rather low.
We found that it is $10^{-9} \mrm{yr}^{-1}$ for comparable-mass IMBH binaries and $10^{-8}$--$10^{-7} \mrm{yr}^{-1}$ for intermediate-mass ratio inspirals (IMRIs) using advanced DPF.

DPF has an ability to detect the gravitational wave background (GWB) with the energy density of $\Omega_\mrm{GW} \ge O(1)$.
However, a stronger bound has already been set from the Big Bang Nucleosynthesis (BBN).
For the observation of the gravity of the Earth, DPF is expected to perform complementary operation compared to other missions such as CHAMP, GRACE and GOCE~\cite{andolisa}.
 (DPF can still observe GWs since geogravity noise dominates other noises only below $f=0.03$Hz.)
However, we do not consider these issues further in this paper and stick to the observation of GWs from IMBH binaries.

This paper is organized as follows.
In Sec.~\ref{sec_IMBH}, we review current observational results of IMBHs and explain the formation mechanisms of IMBHs and their binaries.  
In Sec.~\ref{sec_DPF}, we first review the design and concepts of DPF.
Then, we introduce its noise sensitivity and compare it with the ones of the ground-based detectors.
We derive beam-pattern functions for 1-armed interferometer and explain how to take the effect of DPF motion into account.
In Sec.~\ref{sec_par}, we explain our numerical setups and show the results of both pattern-averaged and Monte Carlo simulations.
We point out possible joint searches with the ground-based detectors in Sec.~\ref{sec_joint} and show the possible constraints on the graviton mass with DPF in Sec.~\ref{sec_graviton}.
In Sec.~\ref{event}, we calculate the expected detection rate of DPF and summarize our work in Sec.~\ref{sec_conclusions}.  
We take $c=G=1$, $H_0=72$km/s/Mpc, $\Omega_m=0.3$ and $\Omega_{\Lambda}=0.7$. 
Throughout this paper, we neglect the eccentricities of the binaries.

\section{Intermediate-Mass Black Holes}
\label{sec_IMBH}

The main target of DPF is IMBH binaries in our Galaxy.
In this section, we briefly review the current observational implications for the existence of IMBHs and the formation mechanisms of IMBHs~\cite{van,colbert,miller2003} and IMBH binaries~\cite{gurkan}.

\subsection{IMBH Observations}

It has been discovered that there exist compact objects with masses larger than 3$\so$ from the careful measurements of the radial velocity of the companion stars.
These objects are considered as the stellar-mass BHs which formed from the core collapses of massive stars.
Also, it is likely that there exist SMBHs at the centers of galaxies.
However the formation process of SMBHs has not been established and one possible way is the collisions of IMBHs with masses $10^2$--$10^4 \so$. 

One type of evidence for the existence of IMBHs is the existence of ULXs whose luminosity exceeds $10^{39}$erg/s, the Eddington luminosity of a $10\so$ BH (see e.g. Ref.~\cite{colbert} for a review on ULXs observations).
This implies that ULXs have masses greater than $10\so$.
On the other hand, they exist several hundred pc away from the galactic centers on average~\cite{colbert1999}.
Since the dynamical friction timescale of SMBHs is smaller than the Hubble time, these BHs are considered to have sunk to the galactic centers by now.
This indicates that the masses of ULXs are smaller than $10^6\so$.
These facts lead to the conclusion that they are likely to be IMBHs.
(There are alternative explanations for ULXs other than IMBHs such as standard stellar-mass BHs with jets or relativistic beamings~\cite{reynolds}.)

Another type of evidence is obtained from the radial mass-to-light ratio $M/L$ which is estimated from the observed profile of the line-of-sight velocity dispersion $\sigma$.
 Gerssen \et~\cite{gerssen} showed that there may exist an IMBH at the center of the galactic globular cluster M15.
Noyola \et~\cite{noyola} observed the surface brightness profile of the globular cluster $\omega$ Centauri, one of the largest and most massive galactic globular clusters.
They claimed clear rise in the velocity dispersion $\sigma$ towards the center
and excluded the ``no BH'' case with more than 99$\%$ confidence level.
However, Anderson and van der Marel~\cite{anderson1,anderson2} analyzed a catalog of 10$^5$ stellar proper motions near the center of $\omega$ Centauri
but did not find any significant rise in the density profile towards the center.
They obtained an upper bound on the mass of IMBH at the center if it exists.   
Recently, Miocchi~\cite{miocchi} analyzed the surface brightness profile of $\omega$ Centauri and claimed that the mass of dark object at the center should lie in the range
$1.3\times 10^4 \so < M < 2.3\times 10^4\so$.
Also, significant core rotations have been observed in the clusters mentioned above and 47 Tuc~\cite{marel,gebhardt2005} which may be signs for the existence of IMBH binaries~\cite{mapelli}.
The possible masses of IMBHs at the centers of the galactic clusters and their distances are summarized in Table~\ref{table-mass1}.

Also, observations of several millisecond pulsars in the galactic globular cluster NGC 6752 indicate that there exists a (1--2)$\times 10^3 \so$ object within the inner 0.08pc of the cluster, assuming that the positive spin derivatives of the pulsars are due to the acceleration by the cluster gravitational potential well~\cite{ferraro}.
Furthermore, Gebhardt \et~\cite{gebhardt2000} and Gerssen \et~\cite{gerssen} found that the rotational speed of the center of M15 is comparable to the velocity dispersion but $N$-body simulation predicts no rotation in the cluster core when there is no massive compact object in the cluster.
Interestingly, this rotation can be explained when there exists a $(20+10^3)\so$ BH binary at the center of the cluster with its orbital separation $a\sim 10^{-3}$pc~\cite{colbert}.
IMBHs are interesting for cluster dynamical evolutions and as the sources for GWs.

\begin{table}[t]
\caption{\label{table-mass1} The distances and possible IMBH masses at the centers of galactic globular clusters.}
\begin{center}
\begin{tabular}{c||c|c}  
 NGC & distance & (total) BH mass  \\ 
No. & (kpc) & ($\so$) \\ \hline
5139 ($\omega$ Cen.) & 4.8~\cite{vandeven} & (3.0--4.75)$\times 10^4$~\cite{noyola} \\
 & & $\leq$1.2$\times 10^4$~\cite{anderson1,anderson2} \\
 & & (1.3--2.3) $\times 10^4$~\cite{miocchi} \\
6388 & 10.0~\cite{harris} & 5.7$\times 10^3$~\cite{lanzoni} \\
6715 (M54) & 26.8~\cite{harris} & 9.4$\times 10^3$~\cite{ibata} \\
6752 & 4.0~\cite{harris} & 2.0$\times 10^3$~\cite{ferraro} \\
7078 (M15) & 10.3~\cite{harris}  & 3.2$\times 10^3$~\cite{gerssen} \\
\end{tabular}
\end{center}
\end{table}

\subsection{IMBH Formation}

There are several IMBH formation mechanisms proposed in globular clusters.
The first one is the repeated hardening of relatively massive BH and 10$\so$ stellar-mass BH binaries via three-body interactions and their mergers due to gravitational radiation.
Since the escape velocity of the typical globular cluster is about 50km/s, massive BHs should have masses larger than 50$\so$ in order to stay in the cluster~\cite{millerhamilton}. 
If initial stars obey the Salpeter mass function, about 10$^{-4}$ of stars have masses larger than 50$\so$, yielding tens of massive BHs in the cluster.
At the center of the cluster, there are comparable amounts of stellar-mass BH and main sequence stars with their number densities $n=10^6 \mrm{pc}^{-3}$.
Within a Hubble time, initially a 50$\so$ BH increases its mass to about $10^3\so$.

The second mechanism is large BHs directly capturing stellar-mass BHs.
If a stellar-mass BH passes near a larger BH sufficiently close, it gets bounded and merges quickly via gravitational radiation.
There is also a formation mechanism proposed by Miller and Hamilton~\cite{millerhamiltonb}, which uses Kozai resonance.
A binary-binary interaction results in a hierarchical triple system.
When certain conditions are realized, the Kozai resonance works, making the eccentricity and the inclination of the inner binary to oscillate.
If the maximum eccentricity becomes around unity, inner binary ends in a quick merger due to the gravitational radiation.

In young clusters with the ages less than a few times 10$^7$ yr, the most massive stars are still on the main sequence.
These will sink to the center of a cluster and a core collapse takes place only among the massive stars.
The resulting high central density leads to the runaway collisions of stars and a very massive star (VMS) formation~\cite{zwart,ebisuzaki,zwart2002,gurkan2004}.
This will eventually collapse to IMBH.
This runaway growth occurs generically in clusters with collapsing times shorter than 3 Myr~\cite{freitag2005}.
IMBHs may also be produced from the gravitational collapse of the population III stars~\cite{colbert}.

\subsection{IMBH Binary Formation}

Binary stars affect the dynamical evolution of globular clusters via dynamical encounters and binary star evolutions. 
Ivanova \et~\cite{ivanova} performed full Monte Carlo N-body simulations and showed that in order to explain the current binary fraction,
the initial fraction needs to be almost 100$\%$.
Then, G$\ddot{\mrm{u}}$rkan \et~\cite{gurkan} performed Monte Carlo simulations of runaway stellar collisions in young dense clusters, 
and found that when the initial fraction is larger than 10$\%$, two VMSs larger than $10^3\so$ form as consequences of the runaway collisions.
First, binary-binary induced runaway collisions take place at off-center of the cluster, producing the first VMS.
Since binary-binary interactions destroy binaries, soon the core binary population is depleted, leading the cluster cores to collapse.
During this phase, the second VMS forms via runaway collisions induced by single-single scatterings.
These VMSs are expected to collapse to IMBHs, eventually forming an IMBH binary.

Another possibility for IMBH binary formation is via cluster mergers.
A large number of young binary clusters have been observed in the Magellanic Clouds~\cite{dieball}.
If they contain IMBHs at their centers, they are expected to form binaries after their mergers.

\section{DECIGO Path Finder}
\label{sec_DPF}

\subsection{Design and Concepts}

DPF is a prototype mission of DECIGO~\cite{ando1,ando2} to test the advanced key technologies of DECIGO such as (i) a precise position measuring system with Fabry-Perot (FP) cavity, (ii) a highly stabilized laser source and (iii) the drag-free control system which shields external forces caused by solar radiation and residual atmosphere.
The weight of the satellite is about 350kg and it will be orbiting the Earth at the Keplerian velocity with an altitude of 500km for 1 yr observation.
It contains 2 mirrors forming a 1-arm interferometer with armlength 30cm.
This interferometer includes a FP cavity with a finesse of about 100 and the laser is emitted from a highly stabilized source with an output power of 100mW at a wavelength of 1030nm.
Currently, the FP cavity has not been tested in space and DPF is expected to have better sensitivity than LPF which uses a Mach-Zender interferometer~\cite{lpf}.
DPF provides new possibilities for a precise position measurement and high-stabilized laser in a space environment. 

Each mirror is placed inside a module called housing.
There are electrostatic-type local sensors on the frame of the housing, which are used for measuring the relative positions of the mirrors and the frame.
The common motion signals of 2 mirrors are fed back to the satellite position using thrusters (drag-free control) while the differential motion signals are sent to the actuators on the frame of the housing to stabilize the FP cavity.
The drag-free controls have been performed by several satellites such as TRIAD-I and Gravity Probe-B satellites.
LPF will operate them at the Lagrange 1 (L1) point where the gravitational environment is stable.
On the other hand, DPF will demonstrate it in an Earth orbit.
This will open a new window for future space missions.

The main GW sources for DPF are IMBH binaries in our Galaxy.
DPF observation is important because observational frequency around 0.1--1Hz cannot be reached by ground-based interferometers and LPF.
Also the development of data analysis technique for DPF has a significant meaning since these data are expected to be more complicated than the one from ground-based interferometers due to the satellite orbital motion and the effects of the Earth.
DPF can also measure the gravity of the Earth with comparable sensitivity to other space missions currently operating and can provide complementary observation to others~\cite{andolisa}.

\begin{figure}[t]
  \centerline{\includegraphics[scale=1.6,clip]{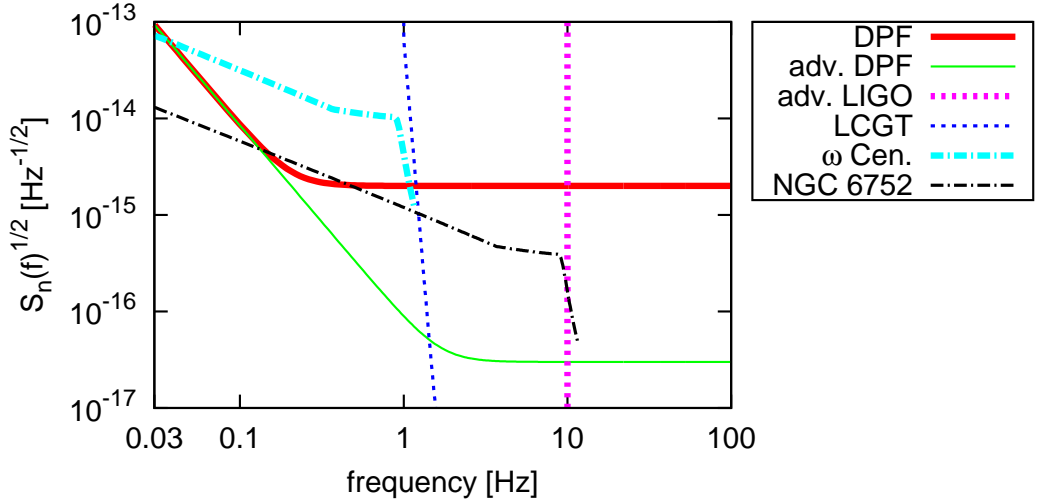} }
 \caption{\label{noise}
The root noise spectral density $\sqrt{S_n(f)}$ of DPF is shown as the (red) thick solid curve, together with the one for adv.~DPF (high frequency part is limited by the shot noise rather than the laser frequency noise) as the (green) thin solid curve.
We also show the adv.~LIGO low frequency cutoff 10Hz as the (magenta) thick dotted line and the one for LCGT down to 1Hz as (blue) thin dotted curve.
The sky-averaged amplitude of a $10^4\so$ equal-mass binary in $\omega$ Centauri is depicted as (light blue) thick dotted-dash curve and the one of a $10^3\so$ equal-mass binary in NGC 6752 is drawn as (black) thin dotted-dashed curve.
Here, we assumed that the observations are carried out from $f=0.03$Hz to mergers.
We also assumed that these BHs have dimensionless effective spin angular momentum $\chi=0.2$. }
\end{figure}

\subsection{Noise Spectrum}

When we assume that the noise $n(t)$ is stationary, the noise spectral density $S_n(f)$ can be defined as
\beq
\left\langle \tilde{n}^{*}(f)\tilde{n}(f')\right\rangle =\delta(f-f')\frac{1}{2}S_n(f), \label{stationary}
\eeq
where $\tilde{n}(f)$ denotes the noise in Fourier domain and  $\left\langle\dots\right\rangle$ represents the expectation value.
The root noise spectral density $\sqrt{S_n(f)}$ of DPF is depicted in Fig.~\ref{noise} as the (red) thick solid curve.
It is given as~\cite{dpf-noise}
\beq
S_n(f)=\frac{1.0\times 10^{-30}}{(2\pi )^4}\lmk \frac{f}{1\mrm{Hz}} \rmk^{-4} +4.0 \times 10^{-30} \ \mrm{Hz}^{-1},
\eeq
where the first term corresponds to the acceleration noises due to the collision of residual gas molecules, the Earth gravity, the magnetic fields of the satellite and the thermal radiation of the housing, while the second term represents the laser frequency noise.
We set the lower frequency cutoff at $f=0.03$Hz due to the Earth gravity.
It may be possible to improve the sensitivity (we call this the ``adv.~DPF'') whose noise spectral density is given as~\cite{andoprivate}
\beq
S_n(f)^{(\mrm{adv})}=\frac{1.0\times 10^{-30}}{(2\pi )^4}\lmk \frac{f}{1\mrm{Hz}} \rmk^{-4} +9.0 \times 10^{-34} \ \mrm{Hz}^{-1}.
\eeq
This time, the high frequency part of the sensitivity is limited by the shot noise. 
Its root noise spectral density is shown as the (green) thin solid curve in Fig.~\ref{noise}.
The (magenta) thick dotted vertical line at $f=10$Hz corresponds to the adv.~LIGO lower cutoff frequency while the (blue) thin dotted vertical one corresponds to the root noise spectral density of LCGT with $f=1$Hz cutoff\footnote{It is likely that adv.~LIGO and adv.~VIRGO also have non-zero sensitivity at $f<10$Hz, but for this frequency rage, LCGT has better sensitivity compared to them. For comparison, Einstein Telescope (ET) is planned to have a sensitivity of $\sqrt{S_n(f)} = 5\times 10^{-21} \mrm{Hz}^{-1/2}$ at $f=1$Hz~\cite{hild}. }.
The latter can be expressed as~\cite{kawamuraprivate}
\beq
S_n(f)^\mrm{LCGT,low}=4.4\times 10^{-25} \lmk \frac{f}{1\mrm{Hz}} \rmk^{-40} + 1.8\times 10^{-32} \lmk \frac{f}{1\mrm{Hz}} \rmk^{-14} .
\eeq

\begin{figure}[t]
  \centerline{\includegraphics[scale=.4,clip]{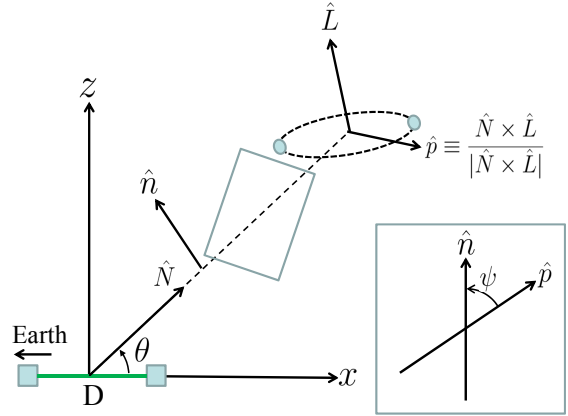} }
 \caption{\label{pol}
We introduce the detector coordinate $\{ x,y,z \}$ with its origin placed at the center of the DPF arm.
The $x$-axis coincides with the detector arm while $z$-axis is orthogonal to this $x$-axis and exists on $\hat{\bm{N}}$-$x$ plane, with $\hat{\bm{N}}$ denoting a unit vector pointing towards the source.
The angle of $\hat{\bm{N}}$ measured from $\hat{\bm{x}}$ is denoted as $\theta$ while the polarization angle $\psi$ is defined as the angle from one of the polarization axes $\hat{\bm{p}}$ to $\hat{\bm{n}}$ which is a unit vector made by projecting $\hat{\bm{z}}$ onto the plane perpendicular to $\hat{\bm{N}}$.
$\hat{\bm{n}}$ is orthogonal to $\hat{\bm{N}}$ and lies on $\hat{\bm{x}}$-$\hat{\bm{z}}$ plane.}
\end{figure}

\subsection{Beam-Pattern Functions and the Effect of Detector Motion}

In this subsection, we first derive the beam-pattern functions for the 1-armed interferometer. 
Let us consider the situation shown in Fig.~\ref{pol}.
The detector arm lies on the $x$-axis and $\hat{\bm{N}}$ denotes the unit vector from the detector to the source.
The $z$-axis lies on the $\hat{\bm{x}}$-$\hat{\bm{N}}$ plane and is orthogonal to the $x$-axis.
$\hat{\bm{L}}$ is the unit orbital angular momentum vector of the binary and $\hat{\bm{n}}$ is a unit vector constructed by projecting $\hat{\bm{z}}$ onto the plane perpendicular to $\hat{\bm{N}}$.
We define $\hat{\bm{p}}\equiv \frac{\hat{\bm{N}}\times \hat{\bm{L}}}{|\hat{\bm{N}}\times \hat{\bm{L}}|}$ which forms one of the polarization principal axes.
The other principal axis $\hat{\bm{q}}$ is defined as $\hat{\bm{q}} \equiv - \frac{\hat{\bm{N}}\times \hat{\bm{p}}}{|\hat{\bm{N}}\times \hat{\bm{p}}|}$.
$\theta$ is the angle of $\hat{\bm{N}}$ measured from $\hat{\bm{x}}$ and the polarization angle $\psi $ is the angle of $\hat{\bm{n}}$ measured from $\hat{\bm{p}}$.

By using the polarization tensors defined as 
\beq
H_{ab}^{+} \equiv p_a p_b - q_a q_b\,, \qquad H_{ab}^{\times} \equiv p_a q_b + q_a p_b\,,
\eeq
the beam-pattern functions for a two-armed interferometer with arm directions $\hat{x}^a$ and $\hat{y}^a$ under $\hat{x}^a \hat{y}_a =0$ can be defined as~\cite{apostolatos,cutler1998}
\beq
F^{+} \equiv \frac{1}{2} H_{ab}^+ (\hat{x}^a \hat{x}^b - \hat{y}^a \hat{y}^b)\,, \qquad  
F^{\times} \equiv \frac{1}{2} H_{ab}^{\times} (\hat{x}^a \hat{x}^b - \hat{y}^a \hat{y}^b)\,.
\eeq
The beam-pattern functions for a one-armed interferometer can be obtained by setting $\hat{y}^a=(0,0,0)$ as
\beqa
F^{+}(\theta,\psi) &=& \frac{1}{2} H_{ab}^+ \hat{x}^a \hat{x}^b = \frac{1}{2}\sin^2\theta \cos 2\psi, \\
F^{\times}(\theta,\psi) &=& \frac{1}{2} H_{ab}^{\times} \hat{x}^a \hat{x}^b = \frac{1}{2} \sin^2\theta \sin 2\psi. \label{beam}
\eeqa
%
Their sky-averaged values of them are $\lla F^{+2} \rra = \lla F^{\times 2} \rra=2/15$.
In Fig.~\ref{noise}, we also show the sky-averaged GW amplitudes of a $10^4\so$ equal-mass IMBH binary in $\omega$ Centauri and a $10^3\so$ equal-mass one in NGC 6752 as the (light blue) thick dotted-dash and the (black) thin dotted-dashed curves, respectively.
We use the phenomenological inspiral-merger-ringdown hybrid waveforms (explained in Sec.~\ref{phenom}) for a circular spin-aligned binary to estimate these amplitudes.
We assume the dimensionless effective spin parameter $\chi$ of these BHs as $\chi=0.2$, where $\chi$ is defined as\footnote{In general, each BH spin has three degrees of freedom. However, since we here assume that each BH spin is (anti-)parallel to the orbital angular momentum, there is only one degree of freedom left (i.e. its magnitude). Furthermore, the dominant contribution of the spin in the waveform appears with the combination shown in Eq.~(\ref{chi-def})~\cite{ajith-spin}. The hybrid waveform that we discuss in Sec.~\ref{phenom} can be accurately described with only one (effective) spin parameter $\chi$~\cite{ajith-spin}. }
\beq
\chi\equiv \frac{1+\delta}{2}\chi_1+ \frac{1-\delta}{2}\chi_2\,.
\label{chi-def}
\eeq
Here, $\delta\equiv (m_1-m_2)/M$ and $\chi_i\equiv S_i/m_i^2$ with $m_i$ and $S_i$ being the mass and the spin angular momentum of $i$-th BH, respectively. 
$\chi$ ranges from -1 to 1, and for the equal spin case, $\chi$ becomes $\chi=\chi_1=\chi_2$.
$\chi =0.2$ is selected as the most probable from Fig.~1 of Ref.~\cite{miller} (see also Ref.~\cite{buliga} for a recent measurement of the spins of the BHs at the centers of globular clusters).
It can be seen that the GW frequency of IMBH binary in $\omega$ Centauri is too low for the ground-based interferometers.
Therefore GW signals of this kind become unique sources for DPF. 
(Higher harmonic signals of ringdown~\cite{bccc} may be detected with LCGT if it is sensitive enough down to 1Hz.)
On the other hand, the one in NGC 6752 can be detected with both DPF and the ground-based ones.
It may be possible to perform joint searches between these detectors (see Sec.~\ref{sec_joint} for more details).

\begin{figure}[t]
  \centerline{\includegraphics[scale=1.4,clip]{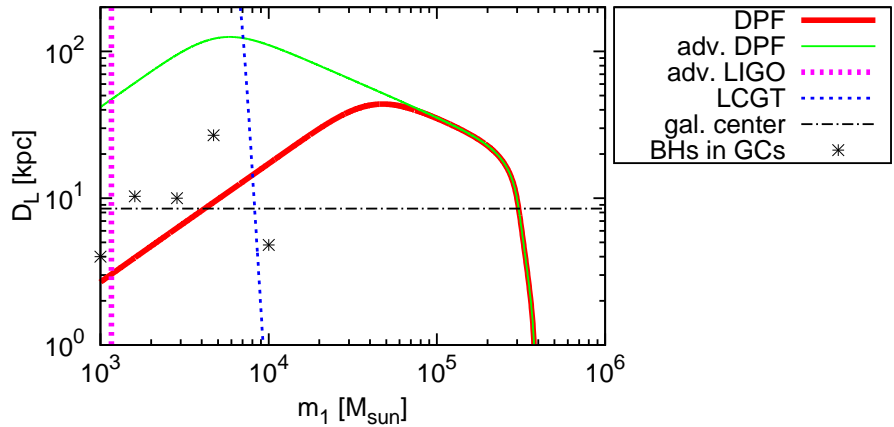} }
 \caption{\label{range_shot}
This figure shows the observable range for each detector with SNR threshold $\rho_\mrm{thr}=5$.
The (red) thick solid curve represents the pattern-averaged one for DPF while the (green) thin solid curve shows the one with adv.~DPF.
Other 2 curves correspond to the same detectors as in Fig.~\ref{noise}. 
We also plot the masses and distances for assumed equal-mass IMBH binaries in the galactic globular clusters shown in Table~\ref{table-mass1}. }
\end{figure}

In Fig.~\ref{range_shot}, we show the (sky-averaged) observable range of DPF with the signal-to-noise ratio (SNR) threshold set to $\rho_\mrm{thr}=5$\footnote{This choice of $\rho_\mrm{thr}$ is rather optimistic and the false alarm rate may not be so small, but we think that $\rho_\mrm{thr}=5$ is the minimum value that we can use to claim the detection (or at least its possibility) of GW signals~\cite{andoprivate}.} as the (red) thick solid curve (see Eq.~(\ref{snr}) for the definition of SNR). 
The (green) thin solid curve corresponds to the one with adv.~DPF.
We also show the one of adv.~LIGO and LCGT with the same curve as in Fig.~\ref{noise}.
The (black) dashed horizontal line at $D_L=8.5$kpc represents the galactic center.
We also plot the possible GW sources in the globular clusters shown in Table~\ref{table-mass1}, assuming that they consist of equal-mass IMBH binaries.

\begin{figure}[t]
  \centerline{\includegraphics[scale=.45,clip]{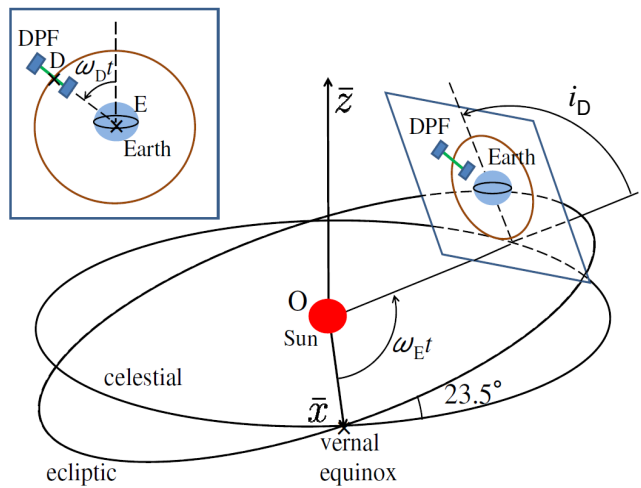} }
 \caption{\label{orbit}
We introduce the celestial coordinate $\{ \bar{x},\bar{y},\bar{z} \}$ with $\bar{x}$-axis pointing the vernal equinox and the $\bar{z}$-axis is orthogonal to the celestial plane and is pointing the celestial North Pole.
DPF orbits the Earth with its plane always facing towards the Sun and its arm pointing the Earth.
It has an inclination angle $i_\mrm{D}=97.3^{\circ}$ against the celestial plane.
The magnified picture of the DPF orbital plane is shown at the top left panel.
The angular velocity of the Earth $\omega_\mrm{E}$ is measured from the $\bar{x}$-axis while the one of the DPF $\omega_\mrm{D}$ is measured from the axis obtained by projecting the $\bar{z}$-axis onto the DPF orbital plane.  }
\end{figure}

Next we introduce the celestial coordinate $\{ \bar{x},\bar{y},\bar{z} \}$ shown in Fig.~\ref{orbit}.
$\bar{x}$-axis points the vernal equinox and $\bar{z}$-axis points the north celestial pole and is orthogonal to the celestial plane.
We express $(\theta,\psi)$ in terms of the source direction $(\bar{\theta}_\mrm{S},\bar{\phi}_\mrm{S})$ and the direction of the orbital angular momentum $(\bar{\theta}_\mrm{L},\bar{\phi}_\mrm{L})$ below.

First, $\theta$ is given as $\cos\theta =\hat{\bm{x}}\cdot \hat{\bm{N}}$ with $\hat{\bm{N}}=(\sin\bar{\theta}_\mrm{S}\cos\bar{\phi}_\mrm{S},\sin\bar{\theta}_\mrm{S}\sin\bar{\phi}_\mrm{S},\cos\bar{\theta}_\mrm{S})$.
(Here, we assumed that distances to sources are much larger than 1AU.)
As shown in Fig.~\ref{range_shot}, DPF orbits the Earth anti-clockwise seen from the Sun, with its arm pointing towards the Earth.
The orbit is solar synchronous and it has orbital inclination of $i_\mrm{D}=97.3^{\circ}$.
At the time $t$, $\hat{\bm{x}}$ is approximately given as
\beq
\hat{\bm{x}}=\lmk \begin{array}{c}
      -\cos\varphi_\mrm{E}(t) \cos\varphi_\mrm{D}(t) \cos i_\mrm{D}-\sin\varphi_\mrm{E}(t)\sin\varphi_\mrm{D}(t)    \\
      -\sin\varphi_\mrm{E}(t) \cos\varphi_\mrm{D}(t) \cos i_\mrm{D}+\cos\varphi_\mrm{E}(t)\sin\varphi_\mrm{D}(t)    \\ 
       \cos\varphi_\mrm{D}(t) \sin i_\mrm{D} \label{ed}
\end{array} \rmk,
\eeq
with $\varphi_\mrm{E}(t)=\omega_\mrm{E} t +\varphi_\mrm{E0}$ and $\varphi_\mrm{D}(t)=\omega_\mrm{D} t +\varphi_\mrm{D0}$.
Here, $\omega_\mrm{E} = 2\pi/1\mathrm{yr}$ and $\omega_\mrm{D}=\sqrt{\frac{M_{\oplus}}{(R_{\oplus}+500\mathrm{km})^3}}$ each represents the angular velocity of the Earth orbiting the Sun and the detector orbiting the Earth, respectively, and $\varphi_\mrm{E0}$ and $\varphi_\mrm{D0}$ denote the position of the detector at $t=0$. 
When deriving Eq.~(\ref{ed}), for simplicity, we assumed that the Earth orbits in the celestial plane rather than in the ecliptic plane.
The difference is negligible as long as the source is situated sufficiently far away from the Sun $(D_L \gg 1\mrm{AU})$. 
Then, $\cos\theta$ becomes
%
\beqa
\fl \cos\theta=&\cos\varphi_\mrm{D}(t) \sin i_\mrm{D} \cos{\bar{\theta}_\mrm{S}} \nonumber \\
\fl & -\left[ \cos\{ \varphi_\mrm{E}(t)-\bar{\phi}_\mrm{S}\} \cos\varphi_\mrm{D}(t) \cos i_\mrm{D}+\sin\{ \varphi_\mrm{E}(t)-\bar{\phi}_\mrm{S}\} \sin\varphi_\mrm{D}(t) \right] \sin{\bar{\theta}_\mrm{S}}. \label{theta}
\eeqa
Next, the polarization angle $\psi$ is estimated as $\cos\psi=\hat{\bm{n}}\cdot \hat{\bm{p}}$.
By using $\hat{\bm{L}}=(\sin\bar{\theta}_\mrm{L}\cos\bar{\phi}_\mrm{L},\sin\bar{\theta}_\mrm{L}\sin\bar{\phi}_\mrm{L},\cos\bar{\theta}_\mrm{L})$ and 
$\hat{\bm{n}}=(\cot\theta) \hat{\bm{N}} -(\sin\theta)^{-1}\hat{\bm{x}}$, it becomes
\beqa
\fl \cos\psi &=\hat{\bm{n}}\cdot \frac{\hat{\bm{N}}\times\hat{\bm{L}}}{|\hat{\bm{N}}\times\hat{\bm{L}}|} \nonumber \\
\fl &=\Bigl[ \cos\bar{\theta}_\mrm{S} \{ -\cos \{ \varphi_\mrm{E}(t)-\bar{\phi}_\mrm{L} \} \sin\varphi_\mrm{D}(t)+\sin \{ \varphi_\mrm{E}(t)-\bar{\phi}_\mrm{L} \} \cos\varphi_\mrm{D}(t)\cos i_\mrm{D} \} \sin\bar{\theta}_\mrm{L}  \nonumber \\
\fl & \quad +\{ \cos \{ \varphi_\mrm{E}(t)-\bar{\phi}_\mrm{S} \} \sin\varphi_\mrm{D}(t) \cos\bar{\theta}_\mrm{L} \nonumber \\
\fl & \quad - \cos\varphi_\mrm{D}(t) \{ \sin \{ \varphi_\mrm{E}(t)-\bar{\phi}_\mrm{S} \} \cos\bar{\theta}_\mrm{L}  \cos i_\mrm{D} 
 + \sin \{ \bar{\phi}_\mrm{L}-\bar{\phi}_\mrm{S} \} \sin\bar{\theta}_\mrm{L}  \sin i_\mrm{D} \} \} \sin\bar{\theta}_\mrm{S} \Bigl] \nonumber \\
 \fl & \quad \Bigl[ (1-\cos^2\theta)\{ 1-(\sin\bar{\theta}_\mrm{S}\sin\bar{\theta}_\mrm{L}\cos(\bar{\phi}_\mrm{L}-\bar{\phi}_\mrm{S})+\cos\bar{\theta}_\mrm{S}\cos\bar{\theta}_\mrm{L})^2 \} \Bigl]^{-1/2}. \label{psi}
\eeqa



\section{Binary Parameter Estimation}
\label{sec_par}

In this section, we perform Fisher analysis to estimate how accurately we can determine the binary parameters if the signals are detected.

\subsection{Fisher Analysis}

If a signal has been detected with a SNR much greater than 1~\cite{vallisneri}, the determination accuracy of the parameter $\theta^i$ is given as~\cite{finn,flanagan}
\beq
\Delta \theta^i = \sqrt{(\Gamma^{-1})_{ii}}, \label{fisherinv}
\eeq
where the Fisher matrix $\Gamma_{ij}$ is defined as
\beq
\Gamma_{ij}\equiv \lmk \frac{\p h}{\p \theta^i} \Big| \frac{\p h}{\p \theta^j} \rmk \label{fisher}
\eeq
with the inner product
\beq
(A|B) \equiv 4 \mathrm{Re}\int ^{\infty}_{0}df \, \frac{\tilde{A}^{*}(f)\tilde{B}(f)}{S_n(f)}. \label{scalar-prod}
\eeq
SNR is defined by using this inner product as
\beq
\rho\equiv \sqrt{(h|h)}. \label{snr}
\eeq


\subsection{Waveform Modeling for IMBH Binaries}  
\label{phenom}

In this paper, we use the phenomenological inspiral-merger-ringdown waveform developed by Ajith \et~\cite{ajith-spin,ajith}.
They basically matched the Taylor T1 post-Newtonian (PN) inspiral waveform~\cite{damour} with the merger and ringdown waveform obtained via numerical simulations.
Then, they parameterized this hybrid waveform with binary parameters.
The Fourier transform of the waveform can be expressed as $\tilde{h}(f) = A(f) e^{i\Psi(f)}$ where the amplitude $A(f)$ and the phase $\Psi(f)$ are given in Eq.~(1) of Ref.~\cite{ajith-spin}.
For the coefficient $C$ that appears in the amplitude, we use~\cite{berti}\footnote{
For the time $t(f)$, we take up to 3.5PN order. 
This can be obtained by integrating the inverse of $df/dt$ given in Arun \et~\cite{arun35} with respect to $f$ and adding the spin-orbit and spin-spin couplings that appear at 1.5PN and 2PN order, respectively (see e.g. Ref.~\cite{berti}).}

\beq
C(t(f)) \equiv \sqrt{\frac{5}{96\pi^{4/3}}}\frac{\mch^{5/6}}{D_L}A_\mrm{pol}(t(f)), \label{c}
\eeq
where $A_\mrm{pol}(t(f))$ is defined as
\beq
 A_\mrm{pol}(t(f)) \equiv \sqrt{(1+\cos^2 i)^2 F^{+}(\theta,\psi)^2+4(\cos i)^2 F^{\times}(\theta,\psi)^2}
\eeq
with $i$ representing the inclination angle of the binary.
The sky-averaged values of the beam-pattern functions are $\lla F^{+2} \rra =\lla F^{\times 2} \rra=2/15$ which yield the sky-averaged value of $C$ as $\lla C \rra =\frac{1}{3\sqrt{10}\pi^{2/3}}\frac{\mch^{5/6}}{D_L}$.

When performing Monte Carlo simulations, we use the waveform
\beq
\tilde{h}(f)=A(f)e^{-i[ \Psi(f)+\varphi_\mrm{pol}(t(f))+\varphi_\mrm{D}(t(f))]},
\eeq
where $\varphi_\mrm{pol}(t(f))$ is the polarization phase given as
\beqa
 \cos[\varphi_{\mathrm{pol}}(t(f))]=&\frac{(1+\cos^2 i)F^{+}(t(f))}{A_{\mathrm{pol}}(t(f))}, \\
 \sin[\varphi_{\mathrm{pol}}(t(f))]=&\frac{2\cos i F^{\times}(t(f))}{A_{\mathrm{pol}}(t(f))}, \label{phipol} 
\eeqa
and $\varphi_\mrm{D}(t(f))$ is the Doppler phase defined as
\beq
\varphi_{D}(t(f))=2\pi f R \sin \bar{\theta}_{\mathrm{S}} \cos[\bar{\phi}(t(f))-\bar{\phi}_{\mathrm{S}}] \label{doppler-phase}
\eeq
with $R\equiv 1$AU.

\subsection{Numerical Setups}

In this subsection, we explain how we performed the Fisher analysis numerically.
We set the cutoff frequencies of DPF as $(f_\mrm{low},f_\mrm{high})=(0.03 \mrm{Hz}, 100 \mrm{Hz})$.
Then, we take the integration range of $(f_{\mathrm{in}},f_{\mathrm{fin}})$ with 
\beq
f_{\mathrm{in}}=\max \bigl\{ f_{\mathrm{low}},  f_\mrm{1yr} \bigr\}, \qquad
f_{\mathrm{fin}}=\min \bigl\{ f_{\mathrm{high}},  f_3 \bigr\}.
\eeq
$f_\mrm{1yr}$ is the frequency at  1 yr before coalescence, which is given as~\cite{flanagan,berti}
\beq
f_\mrm{1yr}=7.38\times 10^{-4} \left[ \left(\frac{\mathcal{M}}{10^4 M_{\odot}}\right)^{-5/8} \right] \mrm{Hz}. \label{fT}
\eeq
Here $\mathcal{M}\equiv M \eta^{3/5}$ is the chirp mass where $M$ is the total mass and $\eta\equiv \mu/M$ is the symmetric mass ratio with the reduced mass $\mu$. 
$f_3$ is the ringdown cutoff frequency given in Ref.~\cite{ajith-spin}.
We impose a prior distribution on $\chi$ as $|\chi| \leq 1$.
Following Berti \et~\cite{berti}, we perform the numerical integration with the Gauss-Legendre routine GAULEG~\cite{numerical}.
This quadrature uses the zero points of the $n$-th Legendre polynomials as the abscissas and the integrand can be calculated exactly up to (2$n$-1)-th order.
For taking the inversion of the Fisher matrix, we use the Gauss-Jordan elimination~\cite{numerical}.
In order to make sure that the inversion has been performed correctly, we first normalize the diagonal components of the Fisher matrix to unity.
Then we perform the inversion and convert it to the inverse of the original Fisher matrix (see Appendix C in Ref.~\cite{kent}).
We multiply the original Fisher matrix with the numerically obtained inverse matrix and see how close the result is to the identity matrix $\delta_{ij}$ to check our inversion scheme.
Among the analyses below, the Monte Carlo simulations for the massive gravity theories (Sec.~\ref{massive-fisher}) is expected to be the most difficult in taking the inverse of the Fisher matrix since it involves largest number of parameters (11 in total). Even in this case, we found that the difference in each component between the numerically calculated and the exact identity matrices is of $\mathcal{O}(10^{-4})$ at most. 
($10^{-4}$ is the criterion used in Berti \textit{et al}.~\cite{berti}.)

\subsection{Pattern-Averaged Analysis}

\begin{figure}
  \begin{center}
    \begin{tabular}{cc}
      \resizebox{75mm}{!}{\includegraphics{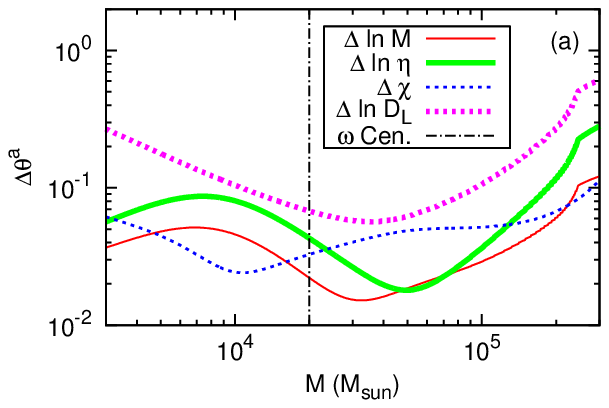}} &
      \resizebox{75mm}{!}{\includegraphics{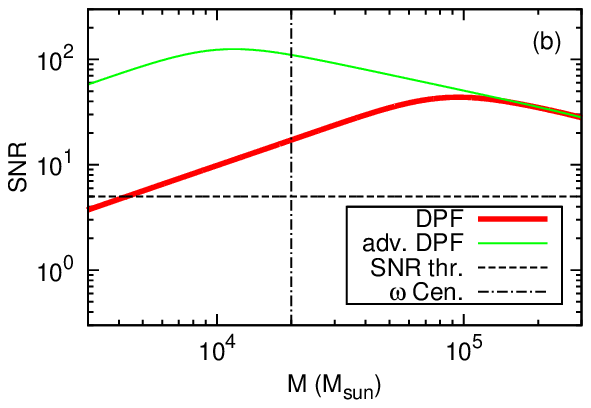}} 
    \end{tabular}
    \caption{(a) DPF Parameter determination accuracies of $\ln M$ (red thin solid), $\ln \eta$ (green thick solid), $\chi$ (blue thin dotted) and $\ln D_L$ (magenta thick dotted) against total mass $M$ for equal-mass binaries of $\chi=0.2$ at $D_L=5$kpc, obtained from pattern-averaged analysis.
The vertical dotted-dashed line represents a possible IMBH binary in $\omega$ Centauri. 
(b) The SNRs for equal-mass binaries of $\chi=0.2$ at $D_L=5$kpc against total mass $M$ with DPF (red thick solid) and with adv.~DPF (green thin solid).
The horizontal dashed line corresponds to the detection threshold $\rho_{\mrm{thr}}=5$. }
    \label{par}
  \end{center}
\end{figure}

In this subsection, we show the results for the pattern-averaged (sky-averaged) analysis.
Here, we have 6 binary parameters in total as
\beq
\bm{\theta}=(\ln M, \ln \eta, \chi, t_0, \phi_0, D_L)
\eeq
with $t_0$ and $\phi_0$ each represents the coalescence time and phase, respectively, and $D_L$ denotes the luminosity distance.
We estimate Fisher matrix by using the phenomenological waveform explained in Sec.~\ref{phenom} and calculate the determination errors of the binary parameters with DPF.
We assumed that IMBH binaries at $D_L=5$kpc contain equal-mass BHs with $\chi=0.2$.
We also set $t_0=\phi_0=0$.
In the panel (a) of Fig.~\ref{par}, we show the estimation errors of $\ln M$ (red thin solid), $\ln \eta$ (green thick solid), $\chi$ (blue thin dotted) and $\ln D_L$ (magenta thick dotted) against the total mass $M$.
The corresponding SNRs are shown in the panel (b) as red thick solid curve.
For an equal-mass IMBH binary in $\omega$ Centauri (dotted-dashed vertical line in the panel (a)), we see that parameters have only a several $\%$ errors.
Therefore DPF may accurately determine the binary parameters if the signals are detected. 
If we use adv.~DPF, roughly speaking, the parameter determination accuracies scale linearly with SNRs shown as a (green) thin solid curve in the panel (b).
For example, the ones for a $(10^4+10^4) \so$ IMBH binary in $\omega$ Centauri improves roughly by a factor 7.
The horizontal dashed line at $\rho=5$ corresponds to the detection threshold.
Since SNRs are not so high, these are only approximate estimations~\cite{cutler-vallisneri}.

\subsection{Monte Carlo Simulations}

Next, following Refs.~\cite{berti,kent,kent2}, we performed Monte Carlo simulations.
This time, we have 10 parameters as
\beq
\bm{\theta}=(\ln M, \ln \eta, \chi, t_0, \phi_0, D_L, \bar{\theta}_\mrm{S},\bar{\phi}_\mrm{S},\bar{\theta}_\mrm{L},\bar{\phi}_\mrm{L}).
\eeq
We consider $10^4\so$ equal-mass IMBH binaries with $\chi=0.2$ at 5 kpc.
For the direction of the sources $(\bar{\theta}_\mrm{S},\bar{\phi}_\mrm{S})$, we take the one of $\omega$ Centauri which has galactic latitude $b$ and longitude $l$ as $(b,l)=(14.97^{\circ},109.10^{\circ})$~\cite{harris}.
This can be converted into $(\bar{\theta}_\mrm{S},\bar{\phi}_\mrm{S})$ by the relations explained in~\ref{galactic}.
Then we randomly generate 10$^4$ sets of $\cos\bar{\theta}_\mrm{L}$ in the range [-1,1] and $\bar{\phi}_\mrm{L}$, $\varphi_\mrm{E0}$ and $\varphi_\mrm{E0}$ in the range [0,$2\pi$].
For each set, we estimate Fisher matrix and obtain the probability distribution of the determination errors at the end.

\begin{figure}[t]
  \centerline{\includegraphics[scale=.8,clip]{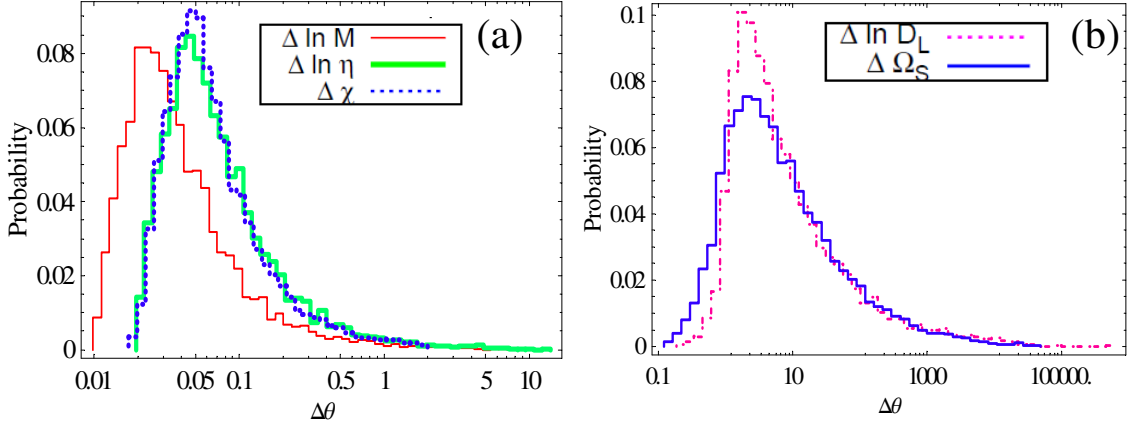} }
 \caption{\label{par_mc}
(a) Histograms showing the parameter determination accuracies of $\ln M$ (red thin solid), $\ln \eta$ (green thick solid) and $\chi$ (blue dotted) using DPF obtained from Monte Carlo simulations.
We assumed $10^4\so$ equal-mass IMBH binaries in $\omega$ Centauri with $\chi =0.2$ and random orientations for the orbital angular momenta.
(b) Histograms showing the parameter determination accuracies of $\ln D_L$ (magenta dotted-dashed) and $\Omega_\mrm{S}$ (blue solid) using DPF.  }
\end{figure}

In Fig.~\ref{par_mc}, we show the probability distributions for the determination errors of the binary parameters using DPF.
In the panel (a), we show the ones for $\ln M$, $\ln \eta$ and $\chi$.
We see that these parameters can be determined with a several $\%$ errors.
In the panel (b), we show the ones for $\ln D_L$ and the angular resolution $\Delta \Omega_\mrm{S}$ with the latter defined as
\beq
\Delta\Omega_S\equiv 2\pi | \sin\bar{\theta}_{\mathrm{S}} | \sqrt{\Sigma_{\bar{\theta}_{\mathrm{S}}\bar{\theta}_{\mathrm{S}}}
                              \Sigma_{\bar{\phi}_{\mathrm{S}}\bar{\phi}_{\mathrm{S}}}-\Sigma^2_{\bar{\theta}_{\mathrm{S}}\bar{\phi}_{\mathrm{S}}}}.
\eeq
Here, $\Sigma_{ij}$ corresponds to the inverse of the Fisher matrix.
It can be seen that these parameters are poorly determined.
The angular resolution can be roughly estimated as
\beq
\Delta\Omega_\mrm{s}\sim \lmk \frac{1}{\rho}\frac{\lambda}{D} \rmk^2 \sim 2.4\lmk \frac{30}{\rho} \rmk^2 \lmk \frac{\lambda}{3\times 10^{10}\mrm{cm}} \rmk^2 \lmk \frac{6.4\times 10^8 \mrm{cm}}{D} \rmk^2,
\eeq
where $\lambda$ is the wavelength of GWs and $D$ is the effective size of the detector which we take as the diameter of Earth.  
This estimate shows a good agreement with our Monte Carlo result.
$\ln D_L$ is determined from the GW amplitude but since the angular resolution is $O(1)$, $\Delta \ln D_L$ also becomes $O(1)$.
Therefore, if we take source directions and orientations into parameters, it is difficult to determine $(\bar{\theta}_\mrm{S},\bar{\phi}_\mrm{S},\bar{\theta}_\mrm{L},\bar{\phi}_\mrm{L})$ and $\ln D_L$.
However, thanks to the weak degeneracies between these parameters and $(\ln M, \ln \eta, \chi)$, the latter set can still be determined fairly accurately.
Since ground-based interferometers are not sensitive for GW signals from IMBH binaries with BH masses $10^4\so$, DPF will perform unique operations for these sources.

\section{Jointed Searches with Ground-Based Interferometers}
\label{sec_joint}

\begin{figure}[t]
  \centerline{\includegraphics[scale=1.5,clip]{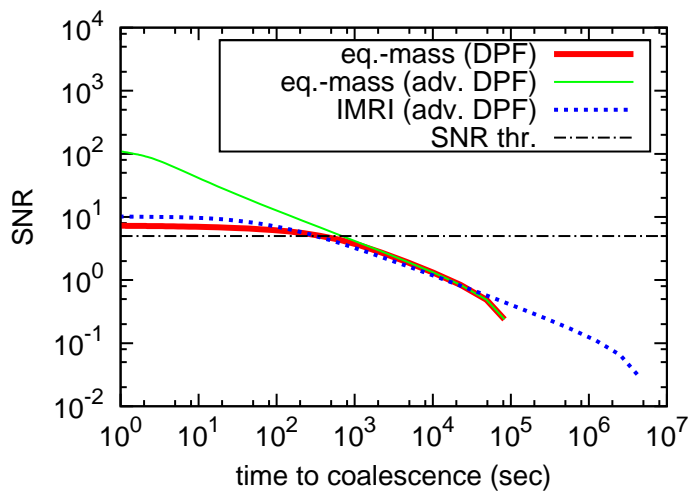} }
 \caption{\label{t}
Accumulated sky-averaged SNRs with DPF (red thick solid) and with adv.~DPF (green thin solid) against the time to coalesce for a $10^3\so$ equal-mass IMBH binary with $\chi=0.2$ in NGC 6752.
We also show the one for a $(10+2\times 10^3)\so$ binary in NGC 6752 with adv.~DPF (blue dotted). 
The horizontal dotted-dashed line corresponds to the detection threshold of  $\rho_\mathrm{thr}=5$.
}
\end{figure}

From Fig.~\ref{noise}, we see that the signals from $10^3\so$ equal-mass IMBH binary in NGC 6752 can be detected with both DPF and the advanced generations of ground-based interferometers.
The (red) thick solid curve in Fig.~\ref{t} shows the accumulated SNR of this signal with DPF against the time to coalesce.
DPF starts detecting this signal at 10$^5$s before coalescence and when $t\approx 4\times10^2$s, the accumulated SNR goes beyond 5. 
When $t=11$s, the GW frequency reaches $f=1$Hz and LCGT may start detecting the signal.
This means that DPF and adv.~DPF may be able to claim detection about 7 mins before coalescence if data analysis can be performed simultaneously\footnote{The number of GW cycles for this signal would be roughly $0.03\mrm{Hz} \times 10^5 \mrm{s} \sim 3 \times 10^3$. As for the data analysis, since the duration of the signal is comparable to (or slightly longer than) the orbital period of the satellite around the Earth (100 mins), we need to calibrate the Doppler-shifted signal and take the change in the position of the satellite into account. However, these can be performed beforehand so that the fundamental part of the data analysis should be almost the same as the usual matched filtering analysis. It would be something similar in between the chirp signal analysis on the ground-based detectors and continuous wave search from pulsars~\cite{andoprivate}. }.
In order to accomplish this goal, DPF needs to send the data to Earth at least every 7 mins.
Then DPF can give alert to the ground-based detectors so as to make sure that they are operating and get ready for the detections.
By combining the data of DPF and the ground-based detectors, we can see the whole history of late inspiral, merger and ringdown of the binary.
(See a related work by Amaro-Seoane and Santamaria~\cite{amaro-joint} on possible joint observations of IMBH binaries with LISA, ET and adv.~LIGO.)
Also, DPF data may help in confirming the actual GW signal detected by the ground-based ones when only the ringdown phase has been detected since they are difficult to distinguish from the noises\footnote{If the galactic IMBH ringdown signals are detected with the ground-based interferometers, the SNRs would be considerably large. 
Therefore, these signals may be confirmed only by taking cross-correlations of the ground-based detectors.}.
If we use adv.~DPF, the time when SNR reaches 5 becomes $t\approx 10^3$s.
In order to give an earlier alert, it is important to reduce the acceleration noise rather than the shot noise.
For example, we found that if we improve the sensitivity of DPF by a factor of two, it would be possible to give alert about 1 hour before coalescence.
For an unequal-mass binary of $(10+2\times 10^3)\so$, the time the signal reaches $\rho=5$ does not change much when we use adv.~DPF (blue dotted). 
 
When only the ringdown signal of a galactic IMBH binary is detected by the ground-based detectors, DPF may help in determining the ringdown efficiency parameter $\epsilon_\mrm{rd}\equiv E_\mrm{GW}/M_\mrm{f}$~\cite{flanaganhughes,bcw} where $E_\mrm{GW}$ is the total energy emitted as radiation and $M_\mrm{f}$ is the mass of the final BH.
Since the ringdown amplitude is proportional to $\sqrt{\epsilon_\mrm{rd}}/D_L$~\cite{bcw}, $\epsilon_\mrm{rd}$ and $D_L$ degenerate.
On the other hand, the direction and orientation of the binary can be determined fairly accurately due to the tremendous amount of SNR with the ground-based detectors.
Then, from the inspiral and merger signals of DPF with SNR $\rho_\mrm{DPF}$, we can determine $D_L$ with an error of $\Delta \ln D_L \approx \rho_\mrm{DPF}^{-1}$.
Using this information, we can determine $\epsilon_\mrm{rd}$ from the ground-based interferometer data with an accuracy roughly the same as $\Delta \ln D_L$.

\section{Probing the Mass of the Graviton}
\label{sec_graviton}


In this section, we also consider the possible constraint on the mass of the graviton $m_g$ from the GW observation with DPF.
Originally, massive gravity theory was proposed by Fierz and Pauli~\cite{fierz} where they simply added Lorentz-invariant mass terms to the Einstein-Hilbert action at the quadratic order.
Nowadays, there are many kinds of massive gravity theories (see e.g. Refs.~\cite{dgp,rubakov2,dubovsky,chamseddine,derham1,derham2}).
Only recently a self-consistent massive gravity theory was proposed~\cite{derham1,derham2} which evades theoretical pathologies like Boulware-Deser ghosts under curved background~\cite{boulware}.
In this paper, we do not stick any specific type of the massive gravity theories.
All we assume is that the graviton has a finite mass $m_g$.

\subsection{Previous Works for the Constraints on the Graviton Mass}

If a graviton has a finite mass, the gravitational potential becomes that of Yukawa type.
This modifies the Kepler's third law and the current solar system experiment places the lower bound on the graviton Compton wavelength $\lambda_g\equiv h/(m_g c)$ as $\lambda_g>2.8 \times 10^{17}$cm~\cite{talmadge}.
A weaker constraint has been obtained from the binary pulsar observations.
Since there are additional gravitational degrees of freedom, the binary evolution due to gravitational radiation changes from that of general relativity and this can be tested from the orbital decay rate of the binary.
By assuming the Fierz-Pauli-type theory, Finn and Sutton obtained the lower bound from PSR B1913+16 and PSR B1534+12 as $\lambda_g>1.6 \times 10^{15}$cm~\cite{sutton}.

In the case of GW observations, the propagation speed of GW changes from the speed of light $c$ and depends on its frequency $f$ when the graviton is massive.
This brings 1PN correction in the phase of GW coming from a binary~\cite{will1998}.
There are many works that calculate the possible constraints on $\lambda_g$ with future GW interferometers such as adv.~LIGO, ET, LISA and DECIGO/BBO~\cite{will1998,yunes,berti,arunwill,stavridis,kent,kent2,keppel,bertigair,huwyler,mirshekari}.
We follow Keppel and Ajith~\cite{keppel} and use the hybrid waveform to estimate the lower bound on $\lambda_g$ with DPF observations of IMBH binaries in our galaxy.

\subsection{Correction in the Gravitational Waveform Phase}

When the graviton has a finite mass $m_g$, its phase velocity is modified from $c$ as~\cite{will1998,kent}
\beq
v_{\mathrm{ph}}^2=\left( 1-\frac{1}{f^2\lambda_g^2}\right)^{-1}.
\eeq
This leads to the correction to the phase of the gravitational waveform in the Fourier domain as~\cite{will1998,keppel}
\beq
\Psi_\mrm{eff}(f)=\Psi(f)-\beta f^{-1}, \label{psig}
\eeq
where $\beta$ is defined as
\beq
\beta \equiv \frac{\pi D}{\lambda_g^2 (1+z)}. \label{beta}
\eeq
Here, the distance $D$ is different from the luminosity distance $D_L$ and it is given as~\cite{will1998}
\beq
D\equiv\frac{1+z}{H_0}\int^z_0\frac{dz'}{(1+z')^2\sqrt{\Omega_{M}(1+z')^3+\Omega_{\Lambda}}}.
\eeq
For galactic IMBH binaries considered in this paper, the difference between $D_L$ and $D$ is very small and the factor $(1+z)$ can be approximated as 1 in the definition of $\beta$ (Eq.~(\ref{beta})).

\subsection{Fisher Analysis and Results}
\label{massive-fisher}

We include $\beta$ to the binary parameters and perform Fisher analysis to estimate how accurately we can determine the binary parameters (especially $\beta$) with DPF observations.
Then, we convert the upper bound on $\beta$ to the lower bound on $\lambda_g$.
First, we give a rough estimate on how strong we can constrain $\lambda_g$ with DPF. 
It is not possible to detect the effect of finite graviton mass if the correction term $\beta f^{-1}$ in Eq.~(\ref{psig}) is smaller than SNR$^{-1}$.
This gives the constraint
\beq
\lambda_g \geq 6.6 \times 10^{17} \mrm{cm} \lmk \frac{D}{5\mrm{kpc}} \rmk^{1/2} \lmk \frac{f}{0.1\mrm{Hz}} \rmk^{-1/2} \lmk \frac{\mrm{SNR}}{30} \rmk^{1/2}. \label{rough_g}
\eeq

\begin{figure}[t]
  \centerline{\includegraphics[scale=1.4,clip]{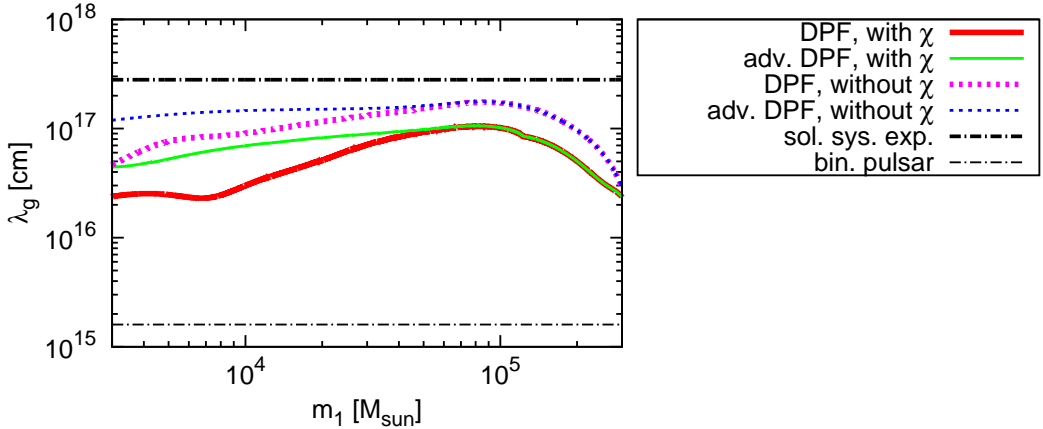} }
 \caption{\label{lambda}
The lower bounds on $\lambda_g$ using DPF with (red thick solid) and without (magenta thick dotted) taking $\chi$ into parameters,  and using adv.~DPF with (green thin solid) and without (blue thin dotted) taking $\chi$ into parameters.
We assumed equal-mass binaries with $\chi=0.2$ at $D_L=5$kpc.
The (black) thin dotted horizontal line at $\lambda_g=2.8\times 10^{17}$cm represents the bound obtained from solar system experiment~\cite{talmadge} while the (black) thick dotted-dashed horizontal line at $\lambda_g=1.6\times 10^{15}$cm shows the one obtained from binary pulsar test~\cite{sutton}.
}
\end{figure}

Next, we estimate the constraint numerically using Fisher analysis.
In Fig.~\ref{lambda}, we show the sky-averaged results for the lower bound on $\lambda_g$ obtained for various BH masses.
We assumed equal-mass binaries with $\chi=0.2$ at $D_L=5$kpc.
The results using DPF are shown in the (red) thick solid curve.
For a $(10^4+10^4)\so$ binary, the constraint becomes $\lambda_g \geq 6\times 10^{16}$cm.
This is weaker than our rough estimate in Eq.~(\ref{rough_g}) and this is due to the degeneracies between $\beta$ and other binary parameters. 
The results using adv.~DPF are shown in the (green) thin solid curve.
We can understand this behaviour by combining Fig.~\ref{noise} and the fact that the constraint on $\lambda_g$ is proportional to SNR$^{-1/2}$ (see Eq.~(\ref{rough_g})). 
In this paper, we only focus on binaries with aligned spins.
When we consider more general cases, there will be precessions which solve the degeneracies between spin and other parameters.
Just to give some idea for the results in these situations, we calculated the constraints on $\lambda_g$ without taking $\chi$ as a variable parameter.
(See Ref.~\cite{stavridis} for the discussion that the constraint on $\lambda_g$ including precession is almost the same as the one without taking spins as variable parameters.)
This corresponds to the situation where the degeneracies between $\lambda_g$ and $\chi$ are completely solved.
The results using DPF and adv.~DPF are shown in the (magenta) thick and the (blue) thin dotted curve, respectively.
The dotted-dashed horizontal line at $\lambda_g =2.8\times10^{17}$cm corresponds to the (static) lower bound obtained from the solar system experiment~\cite{talmadge}.
Although DPF constraint is slightly weaker than this, it is still meaningful since the DPF measures the deviation in the propagation speed of GWs while the solar system experiment measures the deviation in the gravitational constant (or in the Kepler's third law).  
Our results show that DPF can put about 2 orders of magnitude stronger (dynamical) constraint than the one obtained in the weak-field test of the binary pulsar~\cite{sutton} which is shown as the dotted-dashed horizontal line at $\lambda_g =1.6\times10^{15}$cm.
Furthermore, Finn and Sutton~\cite{sutton} assumed Fierz-Pauli-type theory while the constraint obtained here is independent of the specific massive gravity theory.
Since $D_L^{-1}$ in the amplitude cancels with $D$ in $\beta$ when calculating the Fisher matrix (see Eq.~(\ref{rough_g})), the results estimated here are almost independent of $D_L$.
The constraint is stronger for larger mass binaries, hence the constraint becomes stronger if we can also reduce the acceleration noises. 

\begin{figure}[t]
  \centerline{\includegraphics[scale=.45,clip]{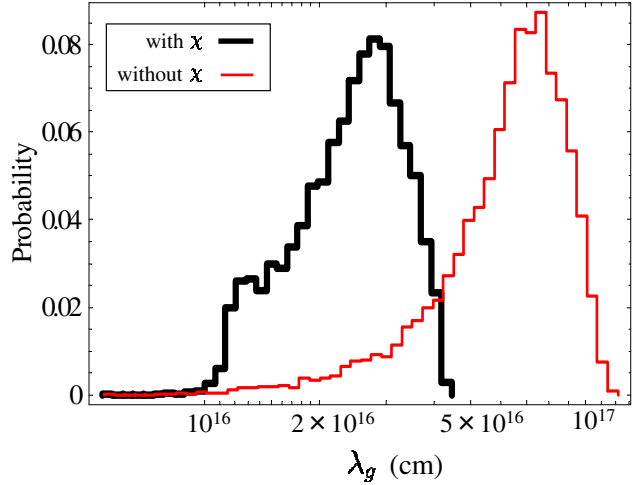} }
 \caption{\label{lambda_mc}
Histograms showing the lower bounds on $\lambda_g$ obtained from Monte Carlo simulations with $10^4\so$ equal-mass IMBH binaries in $\omega$ Centauri with random orientation for the orbital angular momenta.
We show here the results using DPF.
The (black) thick and (red) thin histogram each represents the one with and without taking $\chi$ as a variable parameter, respectively. 
 }
\end{figure}

Next, we performed Monte Carlo simulations for $10^4\so$ equal-mass binaries of $\chi=0.2$ in $\omega$ Centauri. 
The results using DPF are shown in Fig.~\ref{lambda_mc}.
The (black) thick histogram and the (red) thin one each shows the probability distribution for the upper bound on $\lambda_g$ with and without taking $\chi$ as a variable parameter, respectively.
This shows that the constraint becomes slightly weaker compared to the sky-averaged analysis.
However, again, this is much stronger than the one from binary pulsar tests. 
When we use adv.~DPF, these probability distributions shift to larger $\lambda_g$ by roughly $\sqrt{7}=2.6$.

\section{Event Rate Estimations}
\label{event}

\subsection{Mergers of Equal-Mass IMBH Binaries Formed in Galactic Massive Young Clusters}

Currently, more than 10 galactic massive young clusters (GMYCs) have been discovered~\cite{gvaramadze}. 
G$\ddot{\mrm{u}}$rkan \et~\cite{gurkan} performed numerical simulations and found that GMYCs may contain two IMBHs at their centers with BH masses $10^3 \so$, which are likely to form binaries.
After IMBH binary formation, it shrinks due to the dynamical friction with the cluster stars.
The timescale of this process is typically $\leq 10$ Myr which is independent of the local average stellar-mass~\cite{gurkan,fregeau}.
Then, binary shrinks via dynamical encounters with cluster stars, with the timescale of $\leq 1$ Gyr~\cite{fregeau}.
Finally, 2 IMBHs merge due to gravitational radiation within 1 Myr~\cite{gurkan}.

For simplicity, we assume that IMBH binaries are all situated at the galactic center.
From Fig.~\ref{range_shot}, we see that DPF is not sensitive enough to detect GW signals from IMBH binaries in GMYCs.
On the other hand, when we use adv.~DPF, we will be able to see all of the IMBH binaries in GMYCs.
Following Fregeau \et~\cite{fregeau}, we assume that the number of star clusters massive enough to form IMBH binaries equals to the one of globular clusters, and 10$\%$ of them actually produce IMBH binaries.
In our galaxy, there are about 150 globular cluster~\cite{harris}, hence 15 IMBH binaries are expected to be within the reach of DPF.
Since only one IMBH binary is formed over its lifetime for each cluster~\cite{fregeau}, the detection rate of the IMBH binaries in GMYCs with adv.~DPF can be roughly estimated as 
$\nu_\mrm{GMYC}^\mrm{(adv.)} \approx 15/(13.8 \mrm{Gyr}) = 1.1 \times 10^{-9} \mrm{yr}^{-1}$. 



\subsection{Equal-Mass IMBH Mergers in Globular Clusters}

\begin{table}[t]
\caption{\label{table-mass} The distances and velocity dispersions of galactic globular clusters.
Possible masses of IMBHs, if they exit, are obtained from $M-\sigma$ relation~\cite{tremaine}.}
\begin{center}
\begin{tabular}{c||c|c|c}  
 NGC & distance & vel. disp. $\sigma$ & BH mass  \\ 
No. & (kpc)~\cite{harris} &  (km/s)~\cite{dubath} & ($\so$) \\ \hline
104 & 4.5 & 10.0 & 794.7 \\
362 & 8.5 & 6.2 & 116.3 \\ 
1851 & 12.1 & 11.3 & 1299 \\
1904 & 12.9 & 3.9 & 18.04 \\
5272 & 10.4 & 4.8 & 41.57\\
5286 & 11.0 & 8.6 & 433.4 \\
5694 & 34.7 & 6.1 & 108.9 \\
5824 & 32.0 & 11.1 & 1209 \\
5904 & 7.5 & 6.5 & 140.6 \\
5946 & 10.6 & 4.0 & 19.97 \\
6093 & 10.0 & 14.5 & 3539 \\
6266 & 6.9 & 15.4 & 4508 \\
6284 & 15.3 & 6.8 & 168.6 \\
6293 & 8.8 & 8.2 & 357.9 \\
6325 & 8.0 & 6.4 & 132.4 \\
6342 & 8.6 & 5.2 & 57.35 \\
6441 & 11.7 & 19.5 & 11645 \\
6522 & 7.8 & 7.3 & 224.3 \\
6558 & 7.4 & 3.5 & 11.68 \\
6681 & 9.0 & 10.0 & 794.7 \\
7099 & 8.0 & 5.8 & 88.96  \\ 
\end{tabular}
\end{center}
\end{table}

\begin{figure}[t]
  \centerline{\includegraphics[scale=1.3,clip]{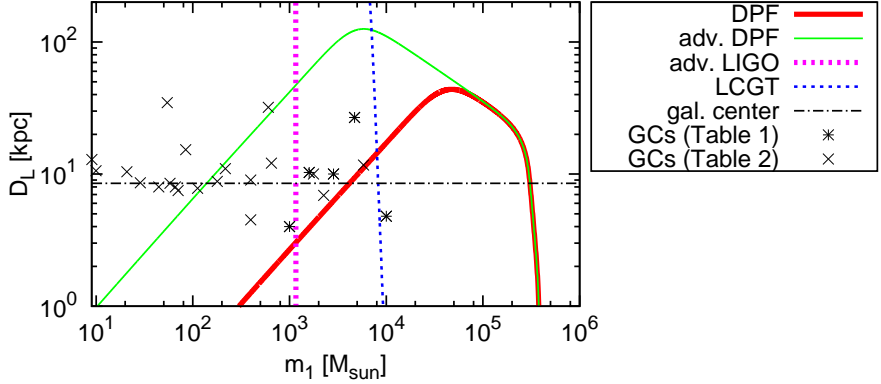} }
 \caption{\label{gc}
Same as Fig.~\ref{range_shot}, but here we also plot possible IMBH binaries shown in Table~\ref{table-mass} as ``$\times$'' with masses assumed from $M-\sigma$ relation. }
\end{figure}

IMBH binaries at the center of galactic globular clusters are also potential GW source candidates for DPF.
Based on Table 6 of Ref.~\cite{dubath}, we select 21 globular clusters out of 150 globular clusters that have been found.
Since the velocity dispersion (denoted as $\sigma_p\mrm{(core)}$) are listed in the table, we apply the $M-\sigma$ formula obtained by Tremaine \et~\cite{tremaine} as
\beq
M=1.35\times 10^8 \lmk \frac{\sigma}{200\mrm{km/s}} \rmk^4 \so,
\eeq
to estimate the mass of the assumed IMBH in each globular cluster.
We assume that this BH is composed of an equal-mass IMBH binary with each BH having mass $M/2$. 
The results are listed in Table~\ref{table-mass}, together with the distance to each globular cluster which is obtained from the catalogue of galactic globular clusters made by Harris (see the URL link shown in Ref.~\cite{harris}). 
We plot IMBH binary of each globular cluster in Fig.~\ref{gc} and count how many of them have SNRs larger than 5 so that they can be detected by DPF.
The meaning of this figure is same as Fig.~\ref{range_shot}. 
We see that about 2 out of 26 globular clusters (5 shown in Table~\ref{table-mass1} + 21 shown in Table~\ref{table-mass}) may contain IMBH binaries that might be detected by DPF on average.
Then, the detection rate of the IMBH binaries in globular clusters $\nu_\mrm{GC}$ is given as 
\beq
\nu_\mrm{GC} \approx 2\times \frac{150}{26}\times\frac{1}{13.8 \ \mrm{Gyr}} = 8.4 \times 10^{-10} \mrm{yr}^{-1}.
\label{rate_GC}
\eeq
Here, we divide the number of globular clusters by the age of the universe for the same reason discussed in the previous subsection.
When we use adv.~DPF, we will be able to see 13 out of 26 IMBH binaries in the galactic globular clusters, making the event rate larger than above by a factor $13/2=6.5$.

\subsection{Intermediate-Mass Ratio Inspirals (IMRIs) with Adv.~DPF}

\begin{figure}[t]
  \centerline{\includegraphics[scale=1.3,clip]{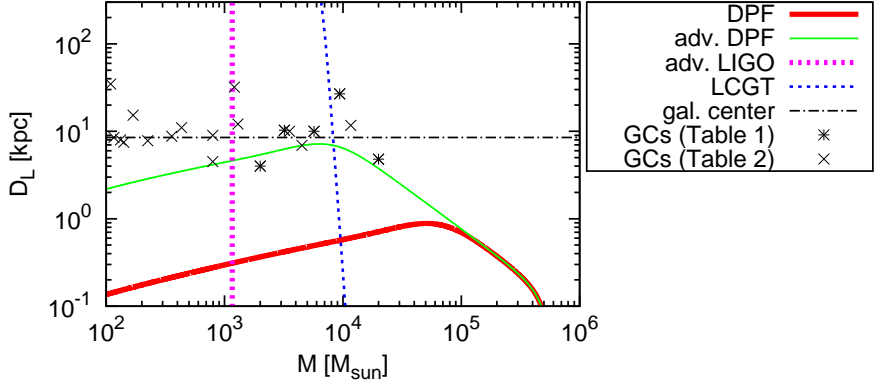} }
 \caption{\label{range_shot_m2_10}
Observable ranges against total mass $M$ with $m_2=10\so$.
The meaning of each curve and plot is same as Fig.~\ref{gc}.}
\end{figure}

Many $10\so$ BHs are expected to form after supernova explosions and they sink to the cluster cores due to the mass segregation.
Therefore at the centers of the clusters, these BHs may have number densities comparable to the main sequence stars, typically $n=10^6$pc$^{-3}$. 
In this subsection, we consider IMBH binaries with intermediate-mass ratio inspirals (IMRIs).
In Fig.~\ref{range_shot_m2_10}, we show observable ranges for DPF (red thick solid) and for adv.~DPF (green thin solid) with $m_2=10\so$.
Meanings of curves and plots are same as in Fig.~\ref{gc}.  
We can see that DPF cannot observe any IMRIs with $m_2=10\so$ but adv.~DPF is sensitive enough to possible IMRIs in some of GCs and may be able to detect an IMRI signal from the galactic center.

\subsubsection{Maximum Merger Rate Per Cluster}
\label{sec_supply}

First, following Ref~\cite{miller}, we will estimate the upper bound for the merger rate per cluster, which is set from (i) the supply limit of smaller BHs, and (ii) the mass of the larger BH.
For case (i), the rate must not exceed $\nu_\mrm{supply}=N_\mrm{BH}/T_\mrm{age}$ where $N_\mrm{BH}$ is the current number of smaller BHs and $T_\mrm{age}$ is the current age of the cluster. 
Otherwise, almost all of the smaller BHs have been consumed by now and $N_\mrm{BH}$ would have been reduced considerably.

For globular clusters with $m_1=10^3\so$, let us assume that the number of stellar-mass BHs with $m_2=10\so$ is $10^3$ (this value is used for NSs with central BH mass $m_1=10^2\so$ in Ref.~\cite{miller}).
Then, the rate for the supply limit becomes
\beq
\nu_\mrm{supply}^\mrm{GC} = \frac{10^3}{13.8 \mrm{Gyr}} = 7.2\times 10^{-8} \mrm{yr}^{-1}.
\eeq
For galactic massive young clusters (GMYCs) with ages less than $3\times 10^7\mrm{yr}$ and with masses larger than $10^4\so$, the number of main sequence stars are at least $10^4\so/0.4\so = 2.5\times10^4$.  
For a multimass King model, the scale height of the smaller BH $m_2$ is smaller compared to the main sequence stars $m_\mrm{ms}$ by a factor $(m_2/m_\mrm{ms})^{1/2}$~\cite{millerhamilton}.
This means that when the number densities of the main sequence stars and the smaller BHs are the same at the cluster core, the number of smaller BHs is smaller by a factor $(m_2/m_\mrm{ms})^{3/2}=125$ (we here  assumed that $m_2=10\so$ and $m_\mrm{ms}=0.4\so$) compared to the main sequence stars. 
Therefore, the number of smaller BHs is roughly $2 \times 10^2 $ and the supply limited rate becomes
\beq
\nu_\mrm{supply}^\mrm{GMYC} = \frac{2\times 10^2}{3\times 10^7 \mrm{yr}} = 6.7\times 10^{-6} \mrm{yr}^{-1}.
\eeq

For case (ii), the merger rate cannot exceed $\nu_\mrm{mass} = (m_1/m_2) T_\mrm{age}^{-1}$ since otherwise, the mass of the larger BH has been increased considerably. For globular clusters with $m_1=10^3\so$, this limit is set as
\beq
\nu_\mrm{mass}^\mrm{GC} = \frac{10^3\so}{10\so} \frac{1}{13.8 \mrm{Gyr}} = 7.2\times 10^{-9} \mrm{yr}^{-1}, 
\eeq
while GMYCs give
\beq
\nu_\mrm{mass}^\mrm{GMYC} = \frac{10^3\so}{10\so} \frac{1}{3\times 10^7 \mrm{yr}} = 3.3\times 10^{-6} \mrm{yr}^{-1}. 
\eeq
In total, the upper bound of the merger rate is set as
\beq
\nu_\mrm{limit} = \min{[\nu_\mrm{supply},\nu_\mrm{mass}]} = \nu_\mrm{mass}.
\label{limit}
\eeq

\begin{figure}[t]
  \centerline{\includegraphics[scale=.45,clip]{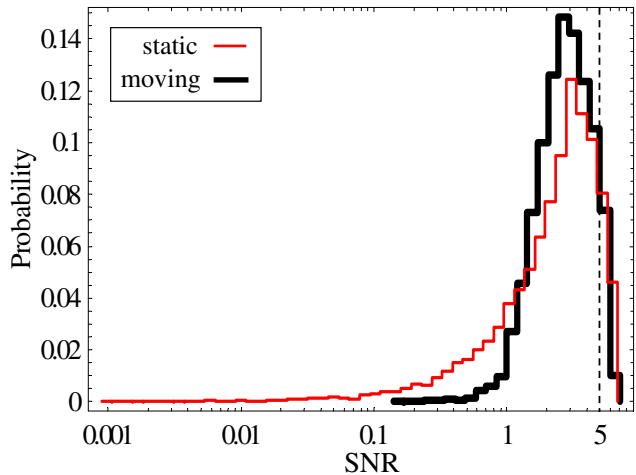} }
 \caption{\label{yc_mc}
The SNR distribution for $10^3\so$ equal-mass IMBH binaries in young clusters at the galactic center. 
We randomly distribute the orbital angular momentum orientations.
The (red) thin and the (black) thick histogram each represents the one for the static and the moving detector, respectively.
The vertical dashed line at $\rho=5$ denotes the SNR threshold.}
\end{figure}

\subsubsection{Merger Rate}

Next, we estimate the merger rate of IMRIs in GCs and GMYCs, with the limit discussed in the previous subsection.
For GCs, we use $\nu_\mrm{limit}^\mrm{GC}=\nu_\mrm{mass}^\mrm{GC}$ as our merger rate per cluster\footnote{Hopman~\cite{hopman} estimated the merger rate of $10\so$ BH into $3\times 10^4\so$ IMBH with Monte Carlo simulations, obtaining $\nu \approx 7\times 10^{-7}$ yr$^{-1}$. 
If we assume that the order does not differ much for smaller IMBH mass cases, this exceeds the limit in Eq.~(\ref{limit}).}.
From Fig.~\ref{range_shot_m2_10}, we see that IMRIs in about 3 out of 26 globular clusters can be seen by adv.~DPF.
This leads to the estimate that it can see IMRIs from about $3\times 150/26=17.3$ globular clusters in total, hence the detection rate of globular clusters in total becomes
\beq
\nu^\mrm{GC,tot}  \lesssim 1.2\times 10^{-7} \mrm{yr}^{-1}.
\label{rate_imp}
\eeq

For GMYCs, from Fig.~\ref{range_shot_m2_10}, we see that adv.~DPF is not sensitive, on average, to an IMRI signal of $(10+10^3)M_\odot$ at galactic center.
However, for some binaries with near optimal orientations of the angular momentum $\bm{L}$, SNRs go beyond $\rho=5$ with DPF. 
Let us consider what fraction of these binaries can be seen by DPF (for simplicity, we here fix the BH mass as $10^3\so$).
From Eq.~(\ref{c}), the angle dependence of SNR squared is given as
\beq
\rho^2 \propto C^2 \propto \sin^4\theta \{  \cos^2 2\psi (1+\cos^2 i)^2 +4 \sin^2 2\psi \cos^2 i \} \equiv X.
\eeq
For a $(10+10^3)\so$ IMRI at $D_L=8.5$kpc with $\rho=5$, $X$ becomes $X=X_0\equiv 3.13$.
Then, the probability that $X$ being greater than $X_0$ is
\beqa
\fl P(X > X_0)  &= \lmk  \frac{1}{2} \rmk^2 \frac{1}{\pi} \int^1_{-1} d\mu_{\theta} \int^1_{-1} d\mu_{i} \int^{\pi}_0 d\psi \nonumber \\
& \quad  \Theta  [ (1-\mu_{\theta}^2)^2 \{ \cos^2 2\psi (1+\mu_i^2)^2+4\sin^2 2\psi \mu_i^2 \}-X_0 ] \nonumber \\
 &=  0.027, \label{probX}
\eeqa
where $\mu_{\theta}\equiv \cos\theta$, $\mu_{i}\equiv \cos i$ and $\Theta [\cdot \cdot \cdot]$ represents the step function.

In order to confirm this analytic estimate, we performed Monte Carlo simulations.
We first consider \textit{static} detectors with $\varphi_\mrm{E}(t)=\varphi_\mrm{E0}$ and $\varphi_\mrm{D}(t)=\varphi_\mrm{D0}$.
Then, we randomly distribute 10$^4$ sets of $\cos\bar{\theta}_\mrm{L}$ in the range [-1,1] and $\bar{\phi}_\mrm{L}$, $\varphi_\mrm{E0}$ and $\varphi_\mrm{D0}$ in the range [0,$2\pi$].
We set the galactic latitude and longitude of the source as $(b,l)=(0,0)$ (galactic center).
We calculate SNR of each set and count the fraction of binaries $P(X > X_0)$ that have SNR greater than 5.
The result of probability distribution function is shown in Fig.~\ref{yc_mc} with (red) thin histogram.
The vertical dashed line corresponds to $\rho=5$.
In this case, we found $P(X > X_0)=0.028$ which shows good agreement with Eq.~(\ref{probX}) of our analytic estimate.
Next, we take the detector motion into account.
We take $\varphi_\mrm{E}(t)=\omega_\mrm{E}t+\varphi_\mrm{E0}$ and $\varphi_\mrm{D}(t)=\omega_\mrm{D}t+\varphi_\mrm{D0}$, and performed the same Monte Carlo simulations.
The (black) thick histogram in Fig.~\ref{yc_mc} is the one obtained with moving detectors.
This time, the fraction becomes $P(X > X_0)=0.017$.
When we take the detector motion into account, SNR approaches to the sky-averaged value and the probability distribution becomes sharper.
This is why $P(X > X_0)$ appeared smaller than the one for the static detector case.
If we manage to improve the sensitivity of adv.~DPF twice, SNRs would be doubled so that   the probability of IMRIs with SNR greater than 5 becomes roughly 50$\%$.
This means that the detection rate would be improved by about 30 times.

According to Ref.~\cite{mapelli2010}, the merger rate due to three-body interaction per cluster becomes $\nu^\mrm{GMYC}=2\times10^{-8}\mrm{yr}^{-1}$.
(For relatively large IMBH masses, the two-body interaction may give comparable contribution compared to the three-body interactions~\cite{miller}.) 
Notice that this value does not exceed the limit we found in Sec.~\ref{sec_supply}.
Gvaramadze \et~\cite{gvaramadze} proposed that there should be 70--100 massive young clusters in the galactic disk in total.
They also suggested that about 50 young massive clusters exist in the galactic center.
Following Ref.~\cite{mapelli2010}, we assume that 75$\%$ of GMYCs contain IMBHs at their centers, meaning that about 50--75 GMYCs may contain IMBHs.
We make a conservative assumption that 50 GMYCs may contain IMBHs.
In this case, the total detection rate becomes 
\beq
\nu^\mrm{GMYC,tot} = 0.017 \times 50 \times 2\times10^{-8}\mrm{yr}^{-1} =1.7\times 10^{-8} \mrm{yr}^{-1}.
\eeq

\section{Conclusions and Discussions}
\label{sec_conclusions}

DPF has an ability to detect IMBH binaries within our galaxy.
Although there is plenty of evidence for the existences of stellar-mass BHs and SMBHs, there is no direct evidence of an IMBH and its existence is still controversial.  
IMBHs may exist at the centers of the galactic globular clusters and the galactic massive young clusters (GMYCs).
For the former cases, globular clusters such as $\omega$ Centauri~\cite{noyola}, M15~\cite{gerssen} and NGC 6752~\cite{ferraro} may harbor IMBHs with masses larger than $10^3\so$.
For the latter, it is suggested that there may exist about 100 GMYCs which contain IMBHs~\cite{gvaramadze}.
A numerical simulation shows that these GMYCs may contain IMBH binaries with 10$^3\so$ BHs~\cite{gurkan}.

If a $10^4\so$ equal-mass IMBH binary exists in $\omega$ Centauri, DPF may perform unique observation since this mass range is too large for the ground-based detectors.
On the other hand, if a $10^3\so$ equal-mass IMBH binary exists in NGC 6752, DPF can see the late inspiral part while LCGT may detect merger and ringdown signals.
Ringdown signal may also be detected by adv.~LIGO and adv.~VIRGO.

We performed both pattern-averaged analysis and Monte Carlo simulations to estimate how accurately we can determine the binary parameters when the signal is detected.
For the latter simulations, we take the motion of DPF into account.
If the one from $\omega$ Centauri is observed, it seems that the total mass can be determined with a 2--3$\%$ accuracy, while the symmetric mass ratio and the effective spin parameter can be determined with 5--6$\%$ accuracies.

We also discussed the possible contributions of DPF to the ground-based GW searches.
First of all, DPF may give alert to the ground-based detectors about 10 mins before coalescence for $10^3\so$ equal-mass IMBH binaries in NGC 6752.
Therefore, the ground-based ones can get ready for their ringdown detections.
Also, DPF data including inspiral and merger signals of IMBH binaries may help in distinguishing the ones from the noises when only ringdown signals have been detected on the ground.
Furthermore, by combining DPF and the ground-based data, we may be able to determine the ringdown efficiency $\epsilon_\mrm{rd}$ which cannot be obtained from the ringdown data alone due to the degeneracy between $\epsilon_\mrm{rd}$ and the distance to the BH.

Also, it may be possible to constrain the graviton Compton wavelength $\lambda_g$ with slightly weaker bound compared to the solar system result in the weak-field regime~\cite{talmadge}.
DPF constraint, independent of the massive gravity theory, is almost 2 orders of magnitude stronger than the binary pulsar tests~\cite{sutton}, which has been obtained by assuming Fierz-Pauli-type theory~\cite{fierz}.   

Unfortunately, the detection rate of DPF is not so high.
For equal-mass IMBH binaries, it is roughly estimated as $10^{-9}$ yr$^{-1}$.
However, when we use adv.~DPF, it has an ability to observe IMBH binaries with $m_2=10\so$ which is a typical mass for the remnant BH of supernova and due to the mass segregation, it is expected that there are plenty of 10$\so$ BHs at the centers of clusters.
Therefore, the detection rate would be larger for these binaries.
For example, the one of $(10+10^3)\so$ in massive young clusters at the galactic center is estimated as $10^{-8}$ yr$^{-1}$--$10^{-7}$ yr$^{-1}$.
Given these values, the prospects of detecting IMBH binaries or IMRIs with IMBHs using DPF or adv.~DPF are not very optimistic.

In this paper, we considered the possibility of reducing the laser frequency noise down to the shot noise.
It is also interesting if we can reduce the acceleration noises as well.
This has more impact on larger mass binaries (like the one in $\omega$ Centauri that we assumed in this paper) and, for instance, the constraint on $\lambda_g$ becomes stronger with the scaling proportional to SNR$^{1/2}$.
Also we will be able to give earlier alert to the ground-based detectors if we can improve the sensitivity on lower frequency part. 
Although the expected detection rate is rather small, the possibility is not zero and since we would be able to obtain interesting science from these binaries (that we discussed in the main part of this paper), we highly recommend to reduce the noise and accomplish at least the sensitivity of adv.~DPF.


\ack
The author gives great thanks to Naoki Seto for useful discussions and checking this manuscript carefully.
The author also thanks Masaki Ando and Takashi Nakamura for valuable comments.
The author thanks Takahiro Tanaka, Seiji Kawamura and Atsushi Nishizawa for fruitful discussions and giving him helpful information.
K.Y. is supported by the Japan Society for the Promotion of Science Grant No. $22 \cdot 900$.
This work is also supported in part by the Grant-in-Aid for the Global COE Program ``The Next Generation of Physics, Spun from Universality and Emergence'' from the MEXT of Japan.

\appendix


\section{Relationship between the Celestial Coordinate and the Galactic Coordinate}
\label{galactic}

\begin{figure}[t]
  \centerline{\includegraphics[scale=.7,clip]{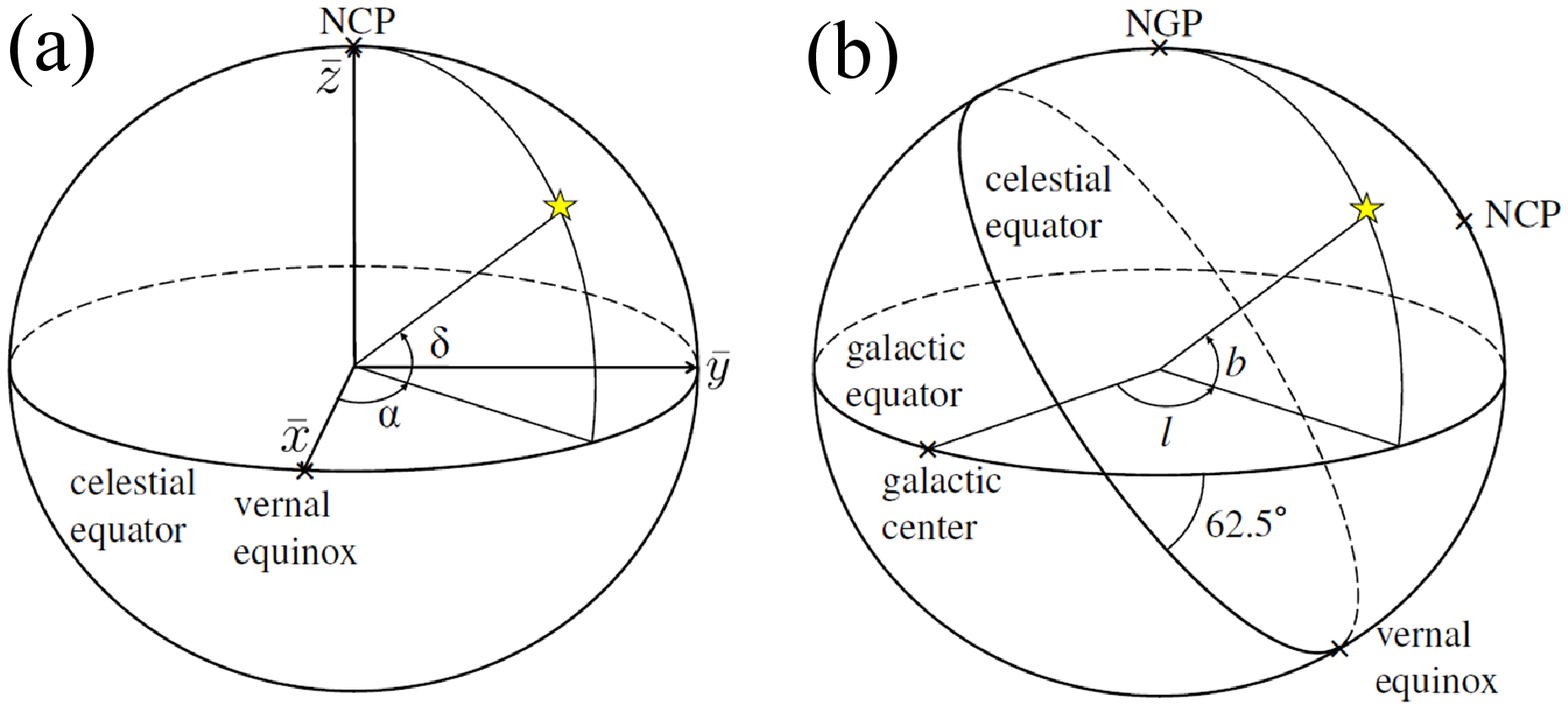} }
 \caption{\label{gal}
(a) The celestial coordinate $\{ \bar{x},\bar{y},\bar{z} \}$ with the right ascension $\alpha$ and the declination $\delta$.
$\bar{x}$ and $\bar{y}$-axes exist on the celestial equator with the former pointing towards the vernal equinox.
$\bar{z}$ points towards the North Celestial Pole (NCP). 
(b) The relation between the celestial plane and the galactic plane with the galactic latitude $b$ and longitude $l$.
$b$ and $l$ are measured from the galactic equator and the galactic center, respectively.
The galactic equator is tilted 62.5$^{\circ}$ from the celestial one.
NGP stands for the North Galactic Pole. }
\end{figure}

We first introduce the right ascension $\alpha$ and the declination $\delta$ shown in the panel (a) of Fig.~\ref{gal}.
The direction of the source $(\bar{\theta}_\mrm{S},\bar{\phi}_\mrm{S})$ can be expressed as $(\bar{\theta}_\mrm{S},\bar{\phi}_\mrm{S}) = (\frac{\pi}{2}-\delta,\alpha)$.
Next, we introduce the galactic coordinate with the galactic latitude $b$ and longitude $l$ shown in the panel (b) of Fig.~\ref{gal}.
$(b,l)$ is related to $(\alpha,\delta)$ as~\cite{binney}
\beqa
\sin\delta = \sin\delta_\mrm{GP} \sin b+ \cos\delta_\mrm{GP} \cos b\cos(l_\mrm{CP}-l), \label{sindel} \\ 
\cos\delta\sin(\alpha-\alpha_\mrm{GP}) = \cos b \sin (l_\mrm{CP}-l), \\
\cos\delta\cos(\alpha-\alpha_\mrm{GP}) = \cos\delta_\mrm{GP} \sin b- \sin\delta_\mrm{GP} \cos b\cos(l_\mrm{CP}-l), \label{cosdel}
\eeqa
where $(\alpha_\mrm{GP},\delta_\mrm{GP})=(192.86^{\circ},27.13^{\circ})$ represents the North Galactic Pole (NGP) and $l_\mrm{CP}=123.93^{\circ}$ is the longitude of the North Celestial Pole (NCP).  
From these equations, $(\bar{\theta}_\mrm{S},\bar{\phi}_\mrm{S})$ can be written as
\beqa
\cos\bar{\theta}_\mrm{S} = & \sin\delta_\mrm{GP} \sin b+ \cos\delta_\mrm{GP} \cos b\cos(l_\mrm{CP}-l), \\
\sin\bar{\theta}_\mrm{S} = & \sqrt{1-\{ \sin\delta_\mrm{GP} \sin b+ \cos\delta_\mrm{GP} \cos b\cos(l_\mrm{CP}-l) \}^2}, \\
\cos\bar{\phi}_\mrm{S} = & \big[-\cos b \sin (l_\mrm{CP}-l) \sin\alpha_\mrm{GP} \nonumber \\
 & + \{ \cos\delta_\mrm{GP} \sin b- \sin\delta_\mrm{GP} \cos b\cos(l_\mrm{CP}-l) \} \cos\alpha_\mrm{GP} \big]/\cos \delta, \\
\sin\bar{\phi}_\mrm{S} =& \big[ \cos b \sin (l_\mrm{CP}-l) \cos\alpha_\mrm{GP}  \nonumber \\
& + \{ \cos\delta_\mrm{GP} \sin b- \sin\delta_\mrm{GP} \cos b\cos(l_\mrm{CP}-l) \} \sin\alpha_\mrm{GP} \big]/\cos \delta.
\eeqa


\vspace{5mm}



\begin{thebibliography}{99}

\bibitem{ligo1}
  G.~M.~Harry  [LIGO Scientific Collaboration],
  Class.\ Quant.\ Grav.\  {\bf 27}, 084006 (2010).

\bibitem{ligo2}
  S.~J.~Waldman  [the LIGO Scientific Collaboration],
  arXiv:1103.2728 [gr-qc].

\bibitem{virgo1}
  T.~Accadia {\it et al.},
  Class.\ Quant.\ Grav.\  {\bf 28}, 114002 (2011).

\bibitem{virgo2}
  T.~Accadia {\it et al.}  [Virgo Collaboration],
  Int.\ J.\ Mod.\ Phys.\  D {\bf 20}, 2075 (2011).

\bibitem{geo1}
  H.~Luck  [LIGO Scientific Collaboration],
  arXiv:1004.0338 [gr-qc].

\bibitem{geo2}
  H.~Grote  [LIGO Scientific Collaboration],
  Class.\ Quant.\ Grav.\  {\bf 27}, 084003 (2010).

\bibitem{lcgt1}
  K.~Kuroda  [LCGT Collaboration],
  Class.\ Quant.\ Grav.\  {\bf 27}, 084004 (2010).

\bibitem{lcgt2}
  K.~Kuroda  [LCGT Collaboration],
  Int.\ J.\ Mod.\ Phys.\  D {\bf 20}, 1755 (2011).

\bibitem{danzmann}
K. Danzmann, Class. Quant. Grav. \textbf{14}, 1399 (1997). 

\bibitem{lisa}
  H.~Araujo {\it et al.},
  J.\ Phys.\ Conf.\ Ser.\  {\bf 314}, 012014 (2011).

\bibitem{allen}
B. Allen, 
Proceedings, \textit{Les Houches, Relativistic Gravitation and Gravitational Radiation} {\bf 373} (1995).




\bibitem{maggiore}
M. Maggiore, Phys. Rept. \textbf{331}, 283 (2000). 

\bibitem{lpf}
M.~Armano {\it et al.},
  Class.\ Quant.\ Grav.\  {\bf 26}, 094001 (2009).

\bibitem{decigo}
N. Seto, S. Kawamura and T. Nakamura, Phys. Rev. Lett. \textbf{87}, 221103 (2001).

\bibitem{kawamura2008}
  S.~Kawamura {\it et al.},
  J.\ Phys.\ Conf.\ Ser.\  {\bf 122}, 012006 (2008).

\bibitem{phinneybbo} 
E. S. Phinney \textit{et al}., \textit{Big Bang Observer Mission Concept Study} (NASA), (2003).

\bibitem{cutlerharms}
C. Cutler and J. Harms, Phys. Rev. \textbf{D73}, 042001 (2006). 


\bibitem{cutlerholz}
C.~Cutler and D.~E.~Holz,
  Phys.\ Rev.\  D {\bf 80}, 104009 (2009).


\bibitem{farmer}
A. J. Farmer and E. S. Phinney, Mon. Not. Roy. Astron. Soc. \textbf{346}, 1197 (2003). 


\bibitem{harms}
J. Harms, C. Mahrdt, M. Otto and M. Priess, Phys. Rev. \textbf{D77}, 123010 (2008).

\bibitem{yagiseto}
K. Yagi and N. Seto, Phys. Rev. \textbf{D83}, 044011 (2011).




\bibitem{nakayama1}
K.~Nakayama, S.~Saito, Y.~Suwa and J.~Yokoyama,
  Phys.\ Rev.\  D {\bf 77}, 124001 (2008).
  
\bibitem{nakayama2}
 K.~Nakayama, S.~Saito, Y.~Suwa and J.~Yokoyama,
  JCAP {\bf 0806}, 020 (2008).
  
\bibitem{gair}
  J.~R.~Gair, I.~Mandel, A.~Sesana and A.~Vecchio,
  Class.\ Quant.\ Grav.\  {\bf 26}, 204009 (2009).

\bibitem{saito1}
  R.~Saito and J.~Yokoyama,
  Phys.\ Rev.\ Lett.\  {\bf 102}, 161101 (2009).
  
\bibitem{saito2}
  R.~Saito and J.~Yokoyama,
  Prog.\ Theor.\ Phys.\  {\bf 123}, 867 (2010).

\bibitem{kent2}
K. Yagi and T. Tanaka, Prog. Theor. Phys. {\bf 123}, 1069 (2010). 

\bibitem{kentbrane}
K. Yagi, N. Tanahashi and T. Tanaka,  Phys. Rev. \textbf{D83}, 084036 (2011).

\bibitem{ando1}
  M.~Ando {\it et al.},
  Class.\ Quant.\ Grav.\  {\bf 26}, 094019 (2009).


\bibitem{ando2}
  M.~Ando {\it et al.},
  Class.\ Quant.\ Grav.\  {\bf 27}, 084010 (2010).


\bibitem{toba}
M.~Ando, K.~Ishidoshiro, K.~Yamamoto, K.~Yagi, W.~Kokuyama, K.~Tsubono and A.~Takamori,
  Phys.\ Rev.\ Lett.\  {\bf 105}, 161101 (2010).

\bibitem{pau1}
P.~Amaro-Seoane, M.~C.~Miller and M.~Freitag, Astrophys.\ J.\ {\bf 692}, L50 (2009).

\bibitem{pau2}
P.~Amaro-Seoane, C.~Eichhorn, E.~K.~Porter and R.~Spurzem, Mon. Not. Roy. Astron. Soc. {\bf 401}, 2268 (2010).

\bibitem{colbert}
M.~C.~Miller and E.~J.~M.~Colbert, Int.\ J.\ Mod.\ Phys.\ {\bf D13}, 1 (2004).


\bibitem{gerssen}
J.~Gerssen, R.~P.~van der Marel, K.~Gebhardt, P.~Guhathakurta, R.~C.~Peterson, C.~Pryor,
Astron.\ J.\  {\bf 124}, 3270 (2002).

\bibitem{noyola}
E.~Noyola, K.~Gebhardt and M.~Bergmann, Astrophys.\ J.\  {\bf 676}, 1008 (2008).

\bibitem{miocchi}
P.~Miocchi,
  Astron.\ Astrophys.\  {\bf 514}, A52 (2010).

\bibitem{anderson1}
J.~Anderson and R.~P.~van der Marel,
  Astrophys.\ J.\  {\bf 710}, 1032 (2010).

\bibitem{anderson2}
R.~P.~van der Marel and J.~Anderson,
  Astrophys.\ J.\  {\bf 710}, 1063 (2010).

\bibitem{ferraro}
 F.~R.~Ferraro,
  Astrophys.\ J.\  {\bf 595}, 179 (2003).



\bibitem{van}
R.~P.~van der Marel,
\textit{The proceedings, Carnegie Observatories Centennial Symposium. 1. Coevolution of Black Holes and Galaxies}, {\bf 20}, 37 (2004). 



\bibitem{pasquato}
  M.~Pasquato,
  ``Croatian Black Hole School 2010 lecture notes on IMBHs in GCs,''
  arXiv:1008.4477 [astro-ph.GA].

\bibitem{gurkan}
  M.~A.~G$\ddot{\mrm{u}}$rkan, J.~M.~Fregeau and F.~A.~Rasio,
  Astrophys.\ J.\  {\bf 640}, L39 (2006).
  

\bibitem{amaro}
P.~Amaro-Seoane and M.~Freitag,
  Astrophys.\ J.\  {\bf 653}, L53 (2006).


\bibitem{ajith-spin}
P.~Ajith {\it et al.},
  Phys. Rev. Lett. \textbf{106}, 241101 (2011).
  
\bibitem{flanaganhughes}
  E.~E.~Flanagan and S.~A.~Hughes,
  Phys.\ Rev.\  D {\bf 57}, 4535 (1998).


\bibitem{bcw}
  E.~Berti, V.~Cardoso and C.~M.~Will,
  Phys.\ Rev.\  D {\bf 73}, 064030 (2006).


\bibitem{keppel}
  D.~Keppel and P.~Ajith,
  Phys. Rev. \textbf{D82}, 122001 (2010).

\bibitem{rubakov}
V. A. Rubakov and P. G. Tinyakov, Phys. Usp. \textbf{51}, 759 (2008). 


\bibitem{hinterbichler}
  K.~Hinterbichler,
  arXiv:1105.3735 [hep-th].


\bibitem{talmadge}
C. Talmadge, J. P. Berthias, R. W. Hellings and E. M. Standish, Phys. Rev. Lett. \textbf{61}, 1159 (1988). 

\bibitem{sutton}
L.~S.~Finn and P.~J.~Sutton,
  Phys.\ Rev.\  D {\bf 65}, 044022 (2002).


\bibitem{andolisa}
A.~Shoda, Y.~Michimura, A.~Araya, Y.~Aso, M.~Ando, W.~Kokuyama, K.~Tsubono and S.~Sato,
Proceedings, \textit{The 8th International LISA Symposium}
(to be published in Journal of Physics: Conference Series (JPCS)).

\bibitem{miller2003}
M.~C.~Miller,
  AIP Conf.\ Proc.\  {\bf 686} (2003) 125.

\bibitem{colbert1999}
E.~J.~M.~Colbert and R.~F.~Mushotzky,
  Astrophys.\ J.\  {\bf 519}, 89 (1999).
  
\bibitem{reynolds}
C.~S.~Reynolds, A.~J.~Loan, A.~C.~Fabian, K.~Makishima, W.~N.~Brandt, T.~Mizuno
Mon. Not. Roy. Astron. Soc. \textbf{286}, 349 (1997).  


\bibitem{marel}
R.~P.~van der Marel, J.~Gerssen, P.~Guhathakurta, R.~C.~Peterson, K.~Gebhardt,
AJ \ {\bf 124}, 3255 (2002).  

\bibitem{gebhardt2005}
  K.~Gebhardt, R.~M.~Rich and L.~C.~Ho,
  Astrophys.\ J.\  {\bf 634}, 1093 (2005).

\bibitem{mapelli}
  M.~Mapelli, M.~Colpi, A.~Possenti and S.~Sigurdsson,
  Mon.\ Not.\ Roy.\ Astron.\ Soc.\  {\bf 364}, 1315 (2005).

\bibitem{lanzoni}
B.~Lanzoni, E.~Dalessandro, F.~R.~Ferraro, P.~Miocchi, E.~Valenti and R.~T.~Rood,
Astrophys.\ J.\  {\bf 668}, L139 (2010). 


\bibitem{ibata}
R.~Ibata {\it et al.},
  Astrophys.\ J.\  {\bf 699}, L169 (2009).

  

\bibitem{vandeven}
G.~van de Ven, R.~C.~E.~van den Bosch, E.~K.~Verolme and P.~T.~de Zeeuw,
  Astron.\ Astrophys.\  {\bf 445}, 513 (2006). 
  
\bibitem{harris}
W.~E.~Harris, AJ \ {\bf 112}, 1487 (1996).    

\bibitem{gebhardt2000}
K.~Gebhardt, C.~Pryor, R.~D.~O'Connell, T.~B.~Williams and J.~E.~Hesser, AJ \ {\bf 119}, 1268 (2000). 


\bibitem{millerhamilton}
M.~C.~Miller and D.~P.~Hamilton,
Mon.\ Not.\ Roy.\ Astron.\ Soc.\  {\bf 330}, 232 (2002).


\bibitem{millerhamiltonb}
M.~C.~Miller and D.~P.~Hamilton,
Astrophys.\ J.\  {\bf 576}, 894 (2002).


\bibitem{zwart}  
S.~F.~Portegies Zwart, J.~Makino, S.~L.~W.~McMillan and P.~Hut, Astron.\ Astrophys.\ {\bf 348}, 117 (1999). 

\bibitem{ebisuzaki}
T.~Ebisuzaki, J.~Makino, T.~G.~Tsuru, Y.~Funato, S.~F.~Portegies Zwart, P.~Hut, S.~McMillan, S.~Matsushita, H.~Matsumoto and R.~Kawabe,  Astrophys.\ J.\  {\bf 562}, L19 (2001).

\bibitem{zwart2002}
S.~F.~Portegies Zwart and S.~L.~W.~McMillan,
  Astrophys.\ J.\  {\bf 576}, 899 (2002).

\bibitem{gurkan2004}
M.~A.~Gurkan, M.~Freitag and F.~A.~Rasio,
  Astrophys.\ J.\  {\bf 604}, 632 (2004).

  
\bibitem{freitag2005}  
M.~Freitag, M.~A.~Gurkan and F.~A.~Rasio,
  Mon.\ Not.\ Roy.\ Astron.\ Soc.\  {\bf 368}, 141 (2006).




\bibitem{ivanova}
 N.~Ivanova, K.~Belczynski, J.~M.~Fregeau and F.~A.~Rasio,
  Mon.\ Not.\ Roy.\ Astron.\ Soc.\  {\bf 358}, 572 (2005).
  
\bibitem{dieball} 
A.~Dieball, H.~M$\ddot{\mrm{u}}$ller and E.~K.~Grebel, A$\&$A\ {\bf 391}, 547 (2002).   

\bibitem{dpf-noise}
M Ando \textit{et al}., Presentation at \textit{9th Edoardo Amaldi Conference on Gravitational Waves} (2011).

\bibitem{andoprivate}
Private communication with Masaki Ando.

\bibitem{hild}
  S.~Hild {\it et al.},
  Class.\ Quant.\ Grav.\  {\bf 28}, 094013 (2011).

\bibitem{kawamuraprivate}
Private communication with Seiji Kawamura.

\bibitem{apostolatos}
T.~A.~Apostolatos, C.~Cutler, G.~J.~Sussman and K.~S.~Thorne,
  Phys.\ Rev.\  D {\bf 49}, 6274 (1994).

\bibitem{cutler1998}
C. Cutler, Phys. Rev. \textbf{D57}, 7089 (1998). 

\bibitem{miller}
M.~C.~Miller,
  Astrophys.\ J.\  {\bf 581}, 438 (2002).

  
\bibitem{buliga}  
S. D. Buliga, V. I. Globina, Yu. N. Gnedin, T. M. Natsvlishvili, M. Yu. Piotrovich and N. A. Shaht, 
arXiv:1108.0056 [astro-ph.CO].
  

\bibitem{bccc}
  E.~Berti, J.~Cardoso, V.~Cardoso and M.~Cavaglia,
  Phys.\ Rev.\  D {\bf 76}, 104044 (2007).
  

\bibitem{vallisneri}
M. Vallisneri, Phys. Rev. \textbf{D77}, 042001 (2008). 
  
\bibitem{finn}
L. S. Finn, Phys. Rev. \textbf{D46}, 5236 (1992). 
  
\bibitem{flanagan}
  C.~Cutler and E.~E.~Flanagan,
  Phys.\ Rev.\  D {\bf 49}, 2658 (1994).

\bibitem{ajith}
P.~Ajith {\it et al.},
  Phys.\ Rev.\  D {\bf 77}, 104017 (2008)
  [Erratum-ibid.\  D {\bf 79}, 129901 (2009)].

\bibitem{damour}
  T.~Damour, B.~R.~Iyer and B.~S.~Sathyaprakash,
  Phys.\ Rev.\  D {\bf 63}, 044023 (2001)
  [Erratum-ibid.\  D {\bf 72}, 029902 (2005)].


\bibitem{berti}
  E.~Berti, A.~Buonanno and C.~M.~Will,
  Phys.\ Rev.\  D {\bf 71}, 084025 (2005).

\bibitem{arun35}
  K.~G.~Arun, B.~R.~Iyer, B.~S.~Sathyaprakash and P.~A.~Sundararajan,
  Phys.\ Rev.\  D {\bf 71}, 084008 (2005).



\bibitem{numerical}
W. H. Press, S. A. Teukolsky, W. T. Vetterling and B. P. Flannery, \textit{Numerical Recipes in Fortran}, Cambridge University Press (1992). 

\bibitem{kent}
K. Yagi and T. Tanaka,  Phys.\ Rev.\  D {\bf 81}, 064008 (2010)
  [Erratum-ibid.\  D {\bf 81}, 109902 (2010)].
 

\bibitem{cutler-vallisneri}
C. Cutler and M. Vallisneri, Phys. Rev. \textbf{D76}, 104018 (2007). 

\bibitem{amaro-joint}
  P.~Amaro-Seoane and L.~Santamaria,
  Astrophys.\ J.\  {\bf 722}, 1197 (2010).


\bibitem{fierz}
M. Fierz and W. Pauli, Proc. Roy. Soc. Lond. \textbf{A173}, 211 (1939). 




\bibitem{dgp}
G. R. Dvali, G. Gabadadze and M. Porrati, Phys. Lett. \textbf{B485}, 208 (2000).  


\bibitem{rubakov2}
V. A. Rubakov, hep-th/0407104. 

\bibitem{dubovsky}
S. L. Dubovsky, JHEP \textbf{0410}, 076 (2004). 

\bibitem{chamseddine}
A.~H.~Chamseddine and V.~Mukhanov,
  JHEP {\bf 1008}, 011 (2010).

\bibitem{derham1}
  C.~de Rham and G.~Gabadadze,
  Phys.\ Rev.\  D {\bf 82}, 044020 (2010).

\bibitem{derham2}
  C.~de Rham, G.~Gabadadze and A.~J.~Tolley,
  Phys.\ Rev.\ Lett.\  {\bf 106}, 231101 (2011).
  
\bibitem{boulware}
  D.~G.~Boulware and S.~Deser,
  Phys.\ Rev.\  D {\bf 6}, 3368 (1972).
  



  
\bibitem{will1998}
C. M. Will, Phys. Rev. \textbf{D57}, 2061 (1998). 
  
\bibitem{yunes}
C. M. Will, and N. Yunes, Class. Quant. Grav. \textbf{21}, 4367 (2004).

\bibitem{arunwill}
K.~G.~Arun and C.~M.~Will,
  Class.\ Quant.\ Grav.\  {\bf 26}, 155002 (2009).

\bibitem{stavridis}
 A.~Stavridis, K.~G.~Arun and C.~M.~Will,
  Phys.\ Rev.\  D {\bf 80}, 067501 (2009).

\bibitem{bertigair}
E.~Berti, J.~Gair and A.~Sesana, arXiv:1107.3528 [gr-qc].

\bibitem{huwyler}
C.~Huwyler, A,~Klein and P.~Jetzer, arXiv:1108.1826 [gr-qc].

\bibitem{mirshekari}
S.~Mirshekari, N.~Yunes and C.~M.~Will, Phys.\ Rev.\ {\bf D85}, 024041 (2012). 

\bibitem{gvaramadze}
V.~V.~Gvaramadze, A.~Gualandris and S.~F.~Portegies~Zwart,
  Mon.\ Not.\ Roy.\ Astron.\ Soc.\  {\bf 385}, 929 (2008).



\bibitem{fregeau}
J.~M.~Fregeau, S.~L.~Larson, M.~C.~Miller, R.~W.~O'Shaughnessy and F.~A.~Rasio,
  Astrophys.\ J.\  {\bf 646}, L135 (2006).

\bibitem{dubath}
P.~Dubath, G.~Meylan and M.~Mayor, Astron.\ Astrophys.\ {\bf 324}, 505 (1997).

\bibitem{tremaine}
S.~Tremaine {\it et al.},
  Astrophys.\ J.\  {\bf 574}, 740 (2002).

\bibitem{hopman}
C. Hopman, 
Class.\ Quant.\ Grav.\ {\bf 26}, 094028 (2009). 

\bibitem{mapelli2010}
M.~Mapelli, C.~Huwyler, L.~Mayer, Ph.~Jetzer and A.~Vecchio,
  Astrophys.\ J.\  {\bf 719}, 987 (2010).

\bibitem{binney}
J.~Binney and M.~Michael,  \textit{Galactic Astronomy}, Princeton University Press (1998). 

\end{thebibliography}
\end{document}